%% file: ms.tex
\documentstyle[11pt,aaspp4,flushrt,psfig2]{article}

\slugcomment{accepted for publication in ApJ}

\lefthead{M.~Kramer et al.}
\righthead{The characteristics of millisecond pulsar emission III.}

\begin{document}

\title{The characteristics of millisecond pulsar emission: \\
 III.~From low to high frequencies}

\author{Michael Kramer\altaffilmark{1,2}, 
Christoph Lange\altaffilmark{2},
Duncan R.~Lorimer\altaffilmark{2,3}
Donald C.~Backer\altaffilmark{1},
Kiriaki M. Xilouris\altaffilmark{3},
Axel Jessner\altaffilmark{2},
Richard Wielebinski\altaffilmark{2}}

\altaffiltext{1}{Astronomy Department, 601 Campbell Hall,
University of California, Berkeley, California 94720, USA}
\altaffiltext{2}{Max-Planck-Institut f\"ur Radioastronomie, Auf dem H\"ugel 69,
53121  Bonn, Germany} 
\altaffiltext{3}{National Astronomy and Ionosphere Center, Arecibo Observatory,
 P.O. Box 995, Arecibo, PR 00613, USA}

\begin{abstract}

In this paper we present the first observations of a large sample of
millisecond pulsars at frequencies of 2.7 GHz ($\lambda11$cm) and 4.9
GHz ($\lambda6$cm). For almost all sources, these represent the first
$\lambda$11cm observations ever.  The new measurements more than
double the number of millisecond pulsars studied at $\lambda$6cm.  Our
new flux measurements extend the known spectra for millisecond pulsars
to the highest frequencies to date.  The coverage of more than a
decade of radio spectrum allows us for the first time to search for
spectral breaks as so often observed for normal pulsars around 1
GHz. The results suggest that, unlike the normal pulsars, millisecond
pulsar spectra can be largely described by a single power law.

We align the observed millisecond pulsar profiles with data from lower
frequencies to search for indications of disturbed magnetic fields and
attempt to resolve questions which were raised in recent literature.
Deviations from a dipolar magnetic field structure are not evident and
absolute timing across the wide frequency range with a single
dispersion measure is possible. We seem to observe mainly unfilled
emission beams, which must originate from a very compact region. The
existence of non-dipolar field components can therefore not be
excluded.

A compact emission region is also suggested by a remarkably constant
profile width or component separation over a very wide frequency
range. This observed difference to the emission properties of normal
pulsars is highly significant. For a few sources, polarization data at
2.7 and 4.9 GHz could also be obtained which indicate that despite the
typically larger degree of polarization at lower frequencies,
millisecond pulsars are weakly polarized or even unpolarized at
frequencies above 3 GHz. The simultaneous decrease in degree of
polarization and the constant profile width thus question proposals
which link de-polarization and decreasing profile width for normal
pulsars to the same propagation effect (i.e.~birefringence).

Comparing the properties of core and conal like profile components to
those of normal pulsars, we find less significant patterns in their
spectral evolution for the population of millisecond pulsars. Hence,
we suggest that core and conal emission may be created by the same emission
process.

Given the small change in profile width, the indicated de-polarization
of the radiation and the possible simple flux density spectra, MSP
emission properties tend to resemble those of normal pulsars only
shifted towards higher frequencies.

\end{abstract}

\section{Introduction}
\label{intro}

Numerous searches following the discovery of the first and still the
fastest rotating neutron star PSR B1937+21 (\cite{bkh+82}) have revealed
about 70 millisecond radio pulsars (e.g.~Manchester 1998).  This sample
has a characteristic age which is on the average two to three orders of
magnitude larger than that of normal pulsars, indicating a different
evolutionary history. Indeed, millisecond pulsars are considered as
being spun-up by mass transfer from a binary companion (\cite{acrs82}).
This evolutionary history distinguishes these recycled pulsars, which we
will here simply call millisecond pulsars, MSPs, from ``normal" pulsars.
Chen \& Ruderman (1993), for instance, suggested that the recycling
process has consequences for the topology of the magnetic field at the
surface of the star which could possibly affect the emission properties
of MSPs. 

One of the first investigations of the emission properties of MSPs was
carried out by \nocite{man92} Manchester (1992). Based upon the small
number of MSPs then known, he concluded that they exhibit properties
similar to normal pulsars. Lorimer et al.~(1995\nocite{lylg95}) found
the spectra of 19 MSPs to be slightly steeper than normal pulsars.  In a
series of papers, we have recently started to study the emission
characteristics of a large sample of MSPs in a systematic way.  Kramer
et al.~(1998, Paper I\nocite{kxl+98}) found that the spectra of normal
and MSPs show great similarities at frequencies up to 1.4 GHz, a
conclusion which was recently confirmed by Toscano et al.~(1998;
hereafter TBMS\nocite{tbms98}) for a large sample of MSPs in the
southern hemisphere at frequencies between 400 and 1600 MHz. In Paper
I we also concluded that MSPs tend to be less luminous compared to
normal pulsars. They are also found to exhibit narrower emission beams
than expected from a period scaling based on data of normal pulsars.
Xilouris et al.~(1998; hereafter Paper II\nocite{xkj+98}) noted that
profiles of MSPs develop much less with frequency and often also in an
unusual way when compared with a canonical behaviour determined from
normal pulsars (Rankin 1983, \cite{lm88}).  Studying a very large sample of
polarization profiles of MSPs for the first time in
Paper II, it was also shown that most MSP polarization position angle
swings appear much flatter than those of normal pulsars. This
observation was recently confirmed by Sallmen (1998\nocite{sal98}) and
Stairs et al.~(1999)\nocite{stc99}. Their results are also in agreement with
Paper II's conclusion that MSP profiles often exhibit a high degree of
polarization which remains almost unchanged up to 1.4 GHz (i.e.~the
highest frequency used in both studies).

Although the data already suggest that the {\em emission mechanism} for
normal and MSPs are essentially identical (see also \cite{jak+98}),
there are prominent differences in their {\em emission properties},
which could be caused by the different evolutionary history (e.g.~the
existence of additional pulse features in a large fraction of MSPs, see
Paper I \& II). The most promising approach to further investigate this
question is a detailed multi-frequency study. In addition to the data
presented in Paper I and II, recent studies provided new high-quality
data for various frequencies below 2 GHz, i.e.~Sallmen
(1998\nocite{sal98}), Stairs et al.~(1999\nocite{stc99}) and TBMS. In
contrast, observations of MSPs at frequencies above 2 GHz are scarce.
The examples which can be found in the literature are summarized in
Table~\ref{oldobs}. Indeed, prior to the observations presented here,
only five MSPs have been detected above 3 GHz, {\it viz}.~PSRs
J0437$-$4715, J1022+1001, J1713+0747, B1855+09 and B1937+21
(see Tab.~\ref{oldobs}). Based on
this small sample, Kramer (1998) \nocite{kra98a} noted that in various
respects the emission properties of MSPs up to 5 GHz tend to mimic those
of normal pulsars at very high frequencies.  A study of this apparent
trend is essential in understanding the physics of the emission
mechanism not only in MSPs but in normal pulsars as well. 

The limited number of MSP detections at high radio frequencies prior to
this work can be mainly attributed to the fact that due to the decrease
in flux density at higher frequencies, large collecting area and wide
observing bandwidths are necessary to obtain sufficient sensitivity to
study the emission properties in detail. Although larger bandwidths are
often available in principle, the detection of short-period pulsars over
large bandwidths is often hampered by the significant amount of
dispersive smearing incurred by incoherent detection schemes.  Recently,
however, a coherent de-disperser has been installed at the 100-m radio
telescope of the MPIfR in Effelsberg, which provides bandwidths of up to
112 MHz. Its combination with a large telescope sensitivity makes this
site ideally suited for observations of MSPs at high frequencies as
shown here. 

In this paper we present the first observations of a large sample of
MSPs at frequencies of 2.7 GHz ($\lambda11$cm) and 4.9 GHz
($\lambda6$cm).  These are the first $\lambda$11cm observations for
almost all MSPs. In this work, we also more than double the number of
detections at $\lambda$6cm. A description of the observing system is
given in Sect.~\ref{observations}.  Our new flux measurements presented
in Sect.~\ref{fluxes} extend the known spectra of MSPs to the highest
frequencies studied so far.  In Sect.~\ref{profiles} we compare our
pulse profiles with low frequency measurements in search for similar
frequency dependencies as found for normal pulsars. After studying the
polarization properties for a number of sources up to 5 GHz in
Sect.~\ref{polsec}, we discuss these results and their implications in
Sect.~\ref{discussion}. 

\section{Observations}

\label{observations}

All data presented in the following for frequencies above 1 GHz are the
result of observations carried out between July 1997 to October 1998
with the 100-m Effelsberg radio telescope operated by the
Max-Planck-Institut f\"ur Radioastronomie, Bonn, Germany.  The observing
system used for 1.4 GHz was almost identical to the one described in
detail in paper I. However, in contrast to Paper I, all profiles
presented here were obtained with the Effelsberg-Berkeley Pulsar
Processor (EBPP), a coherent de-disperser.  The EBPP obtains signals of
two circular polarizations at an
intermediate frequency (IF) of 150 MHz provided by the receiving systems
and converts these to an internal IF of 440 MHz.  A bandpass of, at
maximum, 112 MHz is split into four portions for each of the two
circular polarizations, which are mixed down to baseband.  Each portion
is then sub-divided into eight narrow channels via a set of digital
filters (\cite{bdz+97}).  Their outputs are fed into de-disperser boards
which in turn perform an on-line coherent de-dispersion.  In total 64
output signals (or 128 if full polarization is desired) are detected and
integrated in phase with the topocentric pulse period. Sub-integrations
of typically 180 s are transferred to the controlling computer for
further off-line reduction.  The total bandwidth of the system is
determined by the dispersion smearing of the individual source at the
observing frequency and can be up to 112 MHz. When all four Stokes
parameters are requested, the bandwidth of the EBPP is limited to 28
MHz. 

The 2.7 GHz observations employed a cooled HEMT-receiver operating over
an 80 MHz bandwidth. The system temperature during the observations was
about 40 K on the cold sky with a telescope gain of about ~1.5 K
Jy$^{-1}$, comparable to that at 1.4 GHz.  At 4.9 GHz we used a highly
sensitive HEMT receiver with a system temperature of 25 K on the cold
sky whilst the telescope gain is 1.45 K Jy$^{-1}$.  This receiver
provided either an IF of 750 MHz with a bandwidth of 500 MHz or an IF of
150 MHz with a 200 MHz bandwidth.  For most of the observations, we used
the narrow band configuration fed into the EBPP. For some low DM
pulsars, we also used an incoherent detection system, i.e.~the
Effelsberg Pulsar Observing System (EPOS), connected to a 80-200 MHz
multiplying polarimeter in order to obtain an even wider bandwidth than
provided by the EBPP. A similar set-up was used particularly for polarimetric
observations at 2.7 and 4.9 GHz. For a description of this observing
mode we refer to Kijak et al.~(1997)\nocite{kkwj97}.  The observations
are summarized in Table~\ref{obstab}. 

For comparison with lower frequency data we also present some first
results of Effelsberg observations made at 820 MHz and 1200 MHz with
the EBPP. The HEMT receiver used for these observations is tunable
between 800 and 1300 MHz and has a system noise temperature of about 60 K on
cold sky, providing a 150 MHz IF connected to the EBPP. Due to
the small number of observing sessions, reliable flux measurements are
not available for these particular observations. For all other
frequencies, flux density calibrations were performed in the scheme
detailed in Paper I, i.e.~comparing the pulse energy to the output of a
switch-able noise diode, which itself was calibrated based on observations
of standard flux calibrators, such as 3C123, 3C295 and NGC7027
(\cite{owq+94}). 

The selection of millisecond pulsars to be observed was mainly based on
considerations about integration time available in a given LST range and
the strength of the sources measured at 1.4 GHz. Guided by our
measurements presented in Paper I, we typically selected those sources
which show peak flux densities of $\sim$20 mJy or larger at 1.4 GHz,
i.e.~we did not make any selection based on spectral index.

\section{Flux densities and spectra at high frequencies}

\label{fluxes}

The spectral indices calculated in Paper I were based on only those
high frequency data which were available at that time. The extension
of the frequency coverage to more than a decade of radio spectrum
presented in this paper allows us now for the first time to study the
spectra in search for a steepening in the previously fitted power law
spectra as is often observed for normal pulsars around 1 GHz
(e.g.~\cite{mgj+94}).

Comparing the results of Paper I to those of TBMS, we find that, with
the exceptions of PSRs B1620$-$26, J1643$-$1224 and J2145$-$0750, all of
the derived spectral indices are consistent within the measurement
uncertainties.  For the three aforementioned pulsars, TBMS's new
measurements at frequencies below 1 GHz differ from those values in the
literature used for Paper I.  The low frequency data of TBMS represent a
very large sample of measurements, and thus they should average out
scintillation effects more effectively than previous studies might have
done.  Our flux measurements presented in Paper I agree with
those of TBMS at the overlapping frequencies, and we adopt
TBMS's results for the low frequencies. We also include their 1660 MHz
flux densities in the spectra presented here. For some pulsars, we add
flux densities from Stairs et al.~(1999)\nocite{stc99} and also results
of previously unpublished measurements made at a frequency of 820 MHz at
the 43-m telescope of the National Radio Astronomy
Observatory\footnote{NRAO is operated by Associated Universities, Inc.,
with funding from the National Science Foundation.} in Green Bank.  The
observing setup and calibration procedure of the latter observations,
carried out in April 1997 and 1998, are identical to those described by
Foster et al.~(1991\nocite{ffb91}).  For PSR J2145$-$0750 we also
include a flux measurement made at 102 MHz by Kuzmin \& Losovsky (1996).
The high frequency flux measurements made at Effelsberg typically
represent a number of observing sessions of 30--60 min integration time
for each source. This is sufficient to determine a reliable flux
density, since interstellar scintillation effects are greatly reduced in
amplitude compared to lower frequencies (\cite{mss+96}).  The mean flux
densities, $S_\nu$, are summarized in Table~\ref{spectab}. Resulting
spectra are presented in Fig.~\ref{specfig} (for low frequency data see
also Paper I and references therein). Spectral indices were obtained for
all observed pulsars by a weighted least-squares fit of the power law
expression $S_\nu\propto\nu^\alpha$ and are quoted in
Table~\ref{spectab}. Those sources, which were not detected at 2.7 or
4.9 GHz, but for which interesting upper limits on the flux densities
can be derived, are collected in Table~\ref{upperlim}. 

The first fact which becomes evident from studying Fig.~\ref{specfig} is
that most MSP spectra can be essentially fitted by a {\em
single} power law.  This is an interesting result given the variety seen
in normal pulsar spectra (e.g.~\cite{mgj+94}). The fact that a simple
power law seems to adequately describe MSP spectra well is visible
from the difference in spectral indices, $\Delta \alpha$, derived for
power laws separately fitted to data below and above 1.4 GHz (see
Table~\ref{spectab}) and from the consistency of the spectral indices
presented here and in Paper I. Similarly, for pulsars which are common
both to our and TBMS's study, the spectral indices are in good
agreement, even though our uncertainties are often smaller due to the
larger frequency coverage. The only exception is PSR J1024$-$0724, for
which our observations at 4.9 GHz imply that its spectrum is flatter than
found by TBMS (i.e.~a spectral index of $-1.4\pm0.1$ compared to TBMS's
$-1.70\pm0.12$), although it is also well represented by a simple power
law.

Given that the spectral indices presented here do not differ
significantly from those in Paper I for all studied sources, we do not
expect a change in any of the conclusions drawn in Paper I.  Indeed,
combining the results of this paper, Paper I and those of TBMS, we
still derive a mean spectral index for Galactic field MSPs closer than
1.5 kpc of $-1.76\pm0.14$.  Obviously we have no reason to
believe that the sample of MSPs studied at high frequencies in this
paper comprises a special sub-sample of MSPs.  In fact, although we
did not apply any special criteria to select our sources (apart from a
modest peak flux density at 1.4 GHz), we detected essentially all MSPs
which we tried to observe. The derived upper limits for the
non-detections listed in Tab.~\ref{upperlim} are consistent with the
flux densities expected from power law fits made to the lower frequency
data.

\section{MSP profiles at multiple frequencies}

\label{profiles}

In this section we discuss the frequency development of MSP profiles,
{\em viz.}~their relative arrival time at various frequencies, the
change in profile and component width, the change in component
separation and also the change in relative component amplitudes.  For
this purpose we align profiles according to their barycentric arrival
times at various frequencies using the TEMPO software package
(\cite{tw89}).  This time alignment was performed in a similar way as
for pulsars first described by Foster et al.~(1991)\nocite{ffb91} and later
detailed in Kramer et al.~(1997)\nocite{kxj+97}, i.e.~constructing
noise-free templates by describing the profile as the sum of a number of
Gaussian components (e.g.~\cite{kwj+94}). In the ideal case, this procedure
is perfectly suited for multi-frequency timing measurements. In order to
produce a template for an additional observing frequency only the
relative amplitudes and widths of the components are adjusted while
their locations are kept fixed. As the chosen fiducial point is
therefore the same at all frequencies, a comparison of the pulse
times-of-arrival (TOAs) at different frequencies is possible in an
unbiased way. The TOAs are corrected for dispersion delays according to
their nominal dispersion measures (see Paper I \& II and references
therein) and then compared to a timing model for the particular source.
During this procedure, we usually fit only for the pulse phase, but we
also studied the effects of adjusting the dispersion measure for some
cases.  Only changes within the nominal uncertainties of the known
timing solutions were observed, so that dispersion measures were later
kept fixed to their low frequency values.  The results are presented in
Figs.~\ref{fig0218} -- \ref{fig2145}, where we add low frequency data
taken from the European Pulsar Network (EPN) data base which is freely
available on the World-Wide-Web \footnote{The {\it Internet} address for
this site is http://www.mpifr-bonn.mpg.de/div/pulsar/data/}
(\cite{ljs+98}).  References to the authors who generously contributed
these profiles to the EPN archive, are given in the corresponding figure
captions.  \nocite{stc99} \nocite{snt97} \nocite{bjb+97} \nocite{nst96}
\nocite{fwc93} \nocite{kl96} \nocite{lor94} \nocite{kkwj97}
\nocite{cnst96} The profiles of the new measurements presented here are
also freely available now from the EPN database.  We point out that for
the EPN-profiles timing information was not available. In these cases,
alignments were done visually by bringing the outer edges of the lower
frequency profiles in match with the time-stamped data.  Before
discussing the profile development in general, we make a few notes on
the individual sources.

In total we present profiles for fifteen MSPs.  For five pulsars this is
the first detection around 5 GHz (i.e.~J1012+5307, J1024$-$0719,
J1518+4904, J1643$-$1224 and J1744$-$1134) and also for nine sources at
2.7~GHz (i.e.~J0621+1002, J0751+1808, J1012+5307, J1518+4904,
J1640+2224, J1643$-$1224, J1713+0747, J1744$-$1134 and J2145$-$0750).
Among the eleven MSPs now detected at a frequency as high as 5 GHz, PSR
J0437$-$4715 is the only other source, which is not visible from
Effelsberg. In order to complete this study, we therefore include data
taken from the EPN data base provided by Bell et
al.~(1997)\nocite{bbm+97} and Manchester \& Johnston (1995)\nocite{mj95}
(see Fig.~\ref{fig0437}).  For all sources but PSRs J0751+1807 and
J1640$+$2224 high radio frequency profiles are aligned with data
obtained at lower frequencies. For the latter two MSPs low frequency
data were not available, so that we present their 2.7 GHz profiles
separately in Fig.~\ref{singleprof}.  We also include PSR J0218+4232
here, since our recent measurement at 1.41 GHz represents the highest
frequency for which a profile is available.  Since the number of
observations is too small to exclude scintillation effects for this
source, we do not quote a flux density for PSR J0218+4232, but we align
its profile visually with data from the EPN archive provided by Stairs
et al.~(1999). \nocite{stc99} Finally, for PSRs B1620$-$26 and
J2051$-$0827 we present only spectral information, since the obtained
data are sufficient for reliable flux measurements but the low
signal-to-noise ratio (S/N) of the profiles does not add any additional
knowledge to their profile development with frequency.  All 870,
1220 and 1410-MHz profiles aligned with the high frequency data were
measured in Effelsberg with the EBPP. The 575-MHz profile of PSR J1012+5307
was obtained at Green Bank with the Green-Bank-Berkeley-Pulsar-Processor
(\cite{sal98}), which is identical to the EBPP.

In the following we briefly comment on each individual source
observed. For this discussion we do not arrange the sources in the
usual order of increasing right ascension, but will sort the objects
into three groups, i.e.~isolated, fast spinning and slowly spinning
binary MSPs, respectively.  This is done because in Paper I we have
found some indications that profile properties could be correlated
with the mass of the companion star or, alternatively, with the spin
period of the pulsar. Although we do not find support for these
tentative conclusions of Paper I here, we adopt this scheme for the
merit of future discussions since it has been suggested (e.g.~Chen \&
Ruderman 1993) that the evolutionary history may indeed be reflected
at the emission properties.

\subsection{Isolated MSPs}

In Paper I we confirmed the suggestions made by Bailes et al.~(1997) that
isolated MSPs tend to be less luminous than binary MSPs. Among the
eleven sources detected at 4.9 GHz, three MSPs are isolated objects,
i.e.~PSRs~J1024$-$0719 (Fig.~\ref{fig1024}), J1744$-$1134
(Fig.~\ref{fig1744}) and B1937+21 (Fig.~\ref{fig1937}).  Pulsar B1937+21
is an extraordinarily luminous source, and PSR J1024$-$0719 has a fairly
flat spectrum as discussed above.  By comparison, the spectrum of PSR
J1744$-$1134 is significantly steeper (see Tab.~\ref{spectab}), while
this source also has a smaller luminosity than the average isolated MSP
(Paper I).  We note that the profile of PSR J1744$-$1134
(Fig.~\ref{fig1744}) scarcely changes between 410 and 4848 MHz, although
the presently available S/N prevents a detailed investigation of the
high frequency data. This is consistent with the pre-cursor, which is 
detected at 410, 606 and 1408 MHz,  located at a constant
separation of about 130$^\circ$ prior to the pulse center.
In stark contrast, the profile of PSR J1024$-$0709 changes dramatically
with frequency as visible from the inversion of the component amplitude
ratio. The central component visible at low frequencies weakens
at high frequencies, so that only the leading component remains with a
weak intensity tail at 4.9 GHz (see Fig.~\ref{fig1024}).

The profile development of PSR B1937+21 over the whole spectral range is
more difficult to assess, since at low frequencies the profile is
severely scattered by inhomogeneities in the ISM.  Nevertheless,
inspecting the 320 MHz profile presented by Cordes \& Stinebring
(1984\nocite{cs84}) it seems that the trailing component of the main
pulse, which is clearly visible at 1.4 GHz, is the dominant feature at
this low frequency. It is completely disguised by the scattering tail at
intermediate frequencies, gradually weakens between 1.4, 1.7 and 2.7 GHz
and eventually fades towards high frequencies (cf.~Fig.~\ref{fig1937}).
At the same time the amplitude ratio of main and interpulse increases
gradually with increasing frequency. While the first component of the
main pulse appears unchanged and featureless over the whole spectrum,
the second component gradually weakens relative to the main pulse
(cf.~Cordes \& Stinebring 1984).  This additional component is most
likely not associated with the observed giant pulses of this pulsar
(Jenet et al., in preparation) which appear generally delayed
relative to the position of the main pulse (\cite{bac95}; \cite{cstt96};
Sallmen 1998).

\subsection{Fast rotating binary MSPs ($P<10$ms)}

The steep spectrum of PSR J0218+4232 combined with its large ratio of
dispersion measure to period makes it a difficult source to observe at
high radio frequencies. Consequently, the 1.4-GHz profile shown in
Fig.~\ref{fig0218} is the highest frequency profile available.  Even at 1.4
GHz this pulsar still emits over essentially the full pulse period.
Although wide profiles are also observed in case of normal pulsars
(e.g.~PSR B1831+04; Lyne \& Manchester 1988), the pulse shape of PSR
J0218+4232 and its frequency development are interesting. Indeed,
the central component in the main pulse feature becomes apparently
stronger at higher radio frequencies. At the same time, the weaker pulse
component preceding the main pulse obviously fades with frequency. Stairs
et al.~(1999) present polarization data of this very weakly polarized
pulsar. A determination of the viewing geometry is thus difficult, and
Stairs et al.~(1999) suggest an aligned rotator with a wide beam just
being grazed.

The unusually complex profile of PSR J0437$-$4715 also develops with
frequency (see Fig.~\ref{fig0437} and also Fig.~5 of \cite{nms+97}) in
a manner which is different to what is typically observed for normal
pulsars. In normal pulsars the outer ``conal'' components become
stronger with increasing frequency compared to the central ``core''
component, which eventually even fades away (Rankin 1983, Lyne \&
Manchester 1988). In contrast, in the case of PSR J0437$-$4715 the
amplitude ratio of the central core to its outer conal components
increases from about 2 to 2.5 at 0.4 GHz to more than 13 at 4.6 GHz.
Nevertheless, the outer components are still traceable at this
frequency, which allows us to determine that their separation and the
profile width are apparently unaffected by a change in frequency. We
note however that the position of the core component shifts from a
slightly off-center position to a more centered (earlier) position at
high frequencies. This behaviour is comparable to that of the core
component of PSR B0329+54 (e.g.~\cite{kxj+97}) which also shows
similarities in the overall pulse shape and polarization
characteristics.

Figure~\ref{fig1012} shows a clear profile development for PSR J1012+5307
between 0.6 and 4.9 GHz which, however, does not affect
the profile width or component separation. The trailing component of the
main pulse becomes the dominant feature above 1.4 GHz.  At the same
time, the first interpulse feature separated by about 120$^\circ$ from
the main pulse brightens to about 80\% of the main pulse amplitude at
2.7 GHz and is -- as far as it can be inferred from the given S/N -- as
strong as the main pulse at 4.9 GHz. This trend has already been
noticed in Paper II.

The small misalignment in Fig.~\ref{fig1643} of the profiles of PSR
J1643$-$1224 between 4.9 GHz and the low frequencies is not significant
given the relatively low S/N at 4.9 GHz. Overall, the profile seems to
change very little between low and high frequencies.  Indications of a
possible scattering tail in the 436 MHz profile (\cite{bbm+97}) prevent
a careful comparison with higher frequency data (see also
Sect.~\ref{general}). 

The profile of PSR J1713+0747 (Fig.~\ref{fig1713}) maintains a highly
asymmetric shape throughout the whole spectrum.  The tail-like feature
prominent at low frequencies becomes weaker and almost undetectable at
high frequencies (see also Kijak et al.~1997).  The time alignment
suggests a simultaneous arrival at all frequencies. 

For PSR B1855+09 a distinct profile evolution has already been
demonstrated by Thorsett \& Stinebring (1990) and was extensively
discussed in Paper II. Here, we only emphasise that the separation of
the main pulse components shrinks significantly between 430 MHz and 2380
MHz (see Fig.~4 of \cite{ts90}). This behaviour is comparable to that
for the normal pulsars like PSR B1133+16, which also shows the same
reverse in amplitude ratio in its outer main pulse components as PSR
B1855+09 (\cite{pw92}).  In fact, PSR B1855+09 is one of the very rare
cases where a MSP shows a frequency development in a manner known from
normal pulsars (e.g.~Rankin 1983 and Lyne \& Manchester 1988).
Consistent with that impression is the observation that PSR B1855+09 is
also among the very few MSPs, where the position angle is not flat but
shows an {\sf S}-like swing (cf.~Paper II and references therein). In
this context it is very interesting to compare the profiles observed at
1.4 GHz and at 4.9 GHz (\cite{kkwj97}) as done in Fig.~\ref{fig1855}.
Relative to the main pulse, the interpulse brightens at high frequencies
and bifurcates into two distinct components.  Therefore, in both main
pulse and interpulse the central component has a steeper spectrum than
the outer components.  Applying the knowledge gained from normal pulsars
(as it is apparently justified for this MSP) this frequency behaviour
implies that the observed main and interpulse are created at different
magnetic poles. In other words, PSR B1855+09 is an orthogonal rotator,
which is perfectly consistent with the constant separation of main and
interpulse at all frequencies (see also Foster et al.~1991) and the
observed Shapiro delay (\cite{rt91a}; \cite{ktr94}).

The 2.7-GHz profiles of PSRs J0751+1807 and J1640+2224 shown in
Fig.~\ref{singleprof} are very similar to those presented in Paper I and
II for 1.4 GHz. As described in Paper II, the amplitude of the first
pulse component of PSR J0751+1807 decreases compared to that of the
trailing part, while the central part seems to disappear completely with
increasing frequency. The width of the coherently de-dispersed profile
presented here is consistent with those of the profiles at 430 MHz and
1410 MHz (\cite{lzc95}, Paper I \& II) and the coherently de-dispersed
one at 1410 shown by Sallmen (1998).

\subsection{Slow rotating binary MSPs ($P>10$ms)}

This group of pulsars is distinct from the previous one by having
relatively massive binary companions, i.e.~larger than $0.5$
M${}_{\odot}$. All pulsars discussed below are in fact considered to
form a single family of sources which are spun-up predominantly during a
wind-driven accretion phase, i.e.~producing a non-monotonic random walk
in the angular momentum transferred to the recycled pulsar (in contrast
to disk accretion with monotonic spin-up; see~\cite{vdh94},
\cite{acw99}). 

The first of this family is PSR J0621+1001 for which we present profiles
in Fig.~\ref{fig0621}. The profile remains unchanged between 410 MHz and
2695 MHz and also the component separation and component amplitude
ratio are remarkably constant (see Sect.~\ref{general}). 

When discussing the profile development for PSR J1022+1001, one has to
keep in mind that this source is among the sample of MSPs known to
exhibit profile instabilities. For a detailed discussion we refer the
interested reader to Kramer et al.~(1999)\nocite{kxc+99}. Pulsar
J1022+1001 is another example where a MSP exhibits a distinct but
unusual profile development with frequency. The trailing component, the
dominant feature at 370 MHz, is {\em on average} weaker than the leading
component at 430 MHz, but dominates the profiles up 1.4 GHz
(cf.~Fig.~\ref{fig1022}).  Then, towards high frequencies, this
component weakens again rapidly, being only barely detectable at 4.9 GHz
(see also Kijak et al.~1997).  A similar case is PSR J1518+4904
(Fig.~\ref{fig1518}) where the intensity of the distinct central
component weakens between 370 and 610 MHz, but becomes stronger again at
higher frequencies, until it is the dominant feature at 4848 MHz. At the
same time, the tail-like feature disappears similarly as in the PSR
J1713+0747 profile.  Thus, the leading component appears to have a
maximum in its flux density spectrum around 400 MHz, which would be
consistent with the interpretation that the overall spectrum discussed
before shows a low frequency turn-over. 

Among the strongest profile developments with frequency is the frequency
behaviour of PSR J2145$-$0750: while the leading component is rather
weak at 102 MHz (\cite{kl96}), it becomes the dominant one at higher
frequencies.  The trailing pulse peak, strongest at 102 MHz, weakens at
frequencies of 1.4 GHz and higher, and is hardly visible at 4.8~GHz. The
central pulse component does not seem to develop at all as it remains
the weakest component throughout the whole observed spectrum. Unlike the
normal pulsars, Fig.~\ref{fig2145} indicates that both the profile width
and the component separation hardly change between 100 and 4850 MHz.
Kuzmin \& Losovsky (1996) however claim to observe an increase in
component separation. The fact that the trailing profile
feature\footnote{Note that profiles shown at 102 and 436 MHz have a
limiting time resolution of about 10$^\circ$} consists of two
overlapping components (cf.~Paper I and Sallmen 1998) makes a clear
statement difficult although Sallmen (1998) finds support for a slight
increase in component separation. An inspection of all coherently
de-dispersed profiles available between 410 and 1410 MHz (Sallmen 1998;
Stairs et al.~1999; this work) suggests that neither the component
separation nor the profile width change significantly (see next
section).  Consistently, the prominent pre-cursor does not change its
position as discussed in Paper II. The clear profile change between 102
and 436 MHz raises the question as to whether this is common to all MSPs
at low frequencies. It would be interesting to obtain high quality
measurements of a number of MSPs at these frequencies. 

\subsection{General frequency behaviour}
\label{general}

From the sample of pulse profiles presented in this paper, two obvious
but not necessarily expected facts are evident: first, all profiles
align very well up to the highest frequencies, and second, for some
profiles we observe almost no shape changes at all (e.g.~PSRs J0621+1002
or J1744$-$1134), while a few profiles change dramatically which can be
most easily described by complete reversals in amplitude ratios of
profile components (e.g.~PSRs J0751+1807 or J2145$-$0750).  The
particular examples of PSRs J1518+4904 and J2145$-$0750 demonstrate that
profile changes might occur at frequencies below 400 MHz.  However,
since such data sets are scarce, we can base any discussion only on the
profile developments at frequencies of 400 MHz and above.  At the high
frequency end between 2.7 and 4.9 GHz the observed changes are fairly
small, although some development can be still observed in a few cases
(e.g.~PSR J1022+1001).  In total however, we apparently observe an end
of profile development around 3 GHz.  This frequency seems to be
somewhat lower than the typical frequency for normal pulsars which lies
usually between 5 and 10 GHz (\cite{xkj+96}, \cite{kxj+97}) although it
is not inconsistent (cf.~\cite{srw75}).  Excluding some possible low
frequency profile changes, which may also include scattering effects
imparted on the pulse profile, the data shown here apparently represent
the whole frequency development of MSPs in the radio spectrum. 

In this light, it is worth emphasising that there is an obvious lack
of a change in component separation and MSP profile width. This is a
significant difference to the behaviour of normal pulsars. In order to
demonstrate this behaviour clearly, we followed the example of Foster
et al.~(1991) by measuring the component separation or, if distinct
components were not visible, the profile width (at a 50\% level) as a
function of frequency, $\nu$, for all MSPs discussed in this work (see
Table~\ref{tabsepwidth} and Fig.~\ref{figsepwidth}). For measuring
profile widths, we followed the example of Paper I and measured a 50\%
intensity level with reference to the peaks of the outermost resolved
components. Consequently, reference points used in this work and that
by Foster et al.~(1991) differ for two sources common in both
samples,~PSRs B1620$-$26 and B1855+09, while for PSR B1937+21 both
results are in excellent agreement.  In addition to the shown profiles
we also used data from the literature (e.g.~Paper I), if appropriate
(see Tab.~\ref{tabsepwidth}), and marked coherently (incoherently)
de-dispersed profiles with a circle (triangle), respectively.  We note
that a 50\%-width is not an ideal measure in some cases (e.g.~PSRs
J1643$-$1224 and J1713+0747 where the existence of additional pulse
components at low frequencies obviously influence the measurements),
but offers the possibility for a direct comparison to results
published for normal pulsars. For this purpose, the obtained values
are modelled to an expression $W(\nu)=a_0+a_1\cdot \nu^\gamma$ by
least-squares-fits as done for normal pulsars by Thorsett
(1991)\nocite{tho91a} and Xilouris et al.~(1996)\nocite{xkj+96} (see
Fig.~\ref{figsepwidth}). Obtained indices, $\gamma$, are given in
Tab.~\ref{tabsepwidth}. For most sources, the frequency dependence is
so weak that $a_0$ and $a_1$ are highly correlated. We therefore quote
a width/separation interpolated for 1 GHz instead.  We also emphasise that
although we measure 50\% profile widths in some cases and component
separations in others, we can readily compare the distribution of
indices, $\gamma$, derived for MSPs and normal pulsars, since both
profile width and component separation essentially lead to very
similar frequency dependencies (cf.~\cite{tho91a} and
\cite{xkj+96}). The comparison of histograms is made in
Fig.~\ref{figsepwidthhist}, which demonstrates very clearly the
strikingly different behaviour for both types of objects. A
Kolmogorov-Smirnov test even yields a probability of $10^{-8}$ that
both samples are drawn from the same parent distribution.

\section{Polarization}
\label{polsec}

For the few of the strongest sources at high frequencies or those with
a relative low dispersion measure--period ratio, we obtained
polarization information between 1.7 GHz and 4.9 GHz using the EPOS
and the EBPP (see Sect.~\ref{observations}). Observations are
summarized in Table~\ref{polobstab}, where we quote bandwidth,
dispersion smearing and the degree of linear and total polarization
(i.e.~after adding linear and circular power in quadrature).  The
profiles obtained are essentially unpolarized within the uncertainties
and will be thus shown elsewhere.  This result is interesting as many
MSPs show a modest or even high degree of polarization at lower
frequencies (Paper II, Sallmen 1998).  For comparison we compile data
from Paper II, Manchester \& Johnston (1995)\nocite{mj95}, Navarro et
al.~(1997)\nocite{nms+97}, Sallmen (1998)\nocite{sal98}, Stairs et
al.~(1999)\nocite{stc99}, Thorsett \& Stinebring (1990)\nocite{ts90},
and this work, and plot the degree of polarization for the linearly
polarized intensity as a function of frequency in
Fig.~\ref{polfig}. In Tab.~\ref{tabsepwidth} we list the
de-polarization index, $\epsilon$, which was obtained by fitting the
degree of linear polarization as a function of frequency to a simple
power law, i.e.~$\Pi_L\propto \nu^{\epsilon}$.  Although the sample is
rather small, the data suggest that those sources which are highly
polarized, de-polarize rapidly, while the weakly polarized MSPs
de-polarize even further but apparently with a slower rate.  We stress
that the error bars only reflect instrumental uncertainties which at
2.7 and 4.9 GHz nevertheless include estimates of a possible
de-polarization due to a bandwidth averaging (cf.~\cite{mm98}). This
is important since measuring the polarization of MSPs apparently
involves larger uncertainties compared to studies of normal pulsars,
since we have discovered in Paper II that the polarization
characteristics can show significant temporal variations -- a result
which was later confirmed by Sallmen (1998; cf.~in particular PSRs
J1713+0747 and J2145$-$0750) and Stairs et al.~(1999). As a result,
the degree of polarization can vary significantly for some sources,
e.g.~occasionally a profile is observed which is much more strongly
polarized than the average one. Single pulse studies are needed to
address this question further, but we believe that an average of many
independent observation leads to a representative value.

\section{Discussion}
\label{discussion}

We have demonstrated that the time alignment of MSP profiles presented
in our work for frequencies between 0.9
and 4.9 GHz\footnote{0.6 and 4.9 GHz for PSR J1012+5307, respectively.} 
 is possible without
taking into account timing irregularities caused by effects such as
aberration, magnetic field sweep-back, magnetic multi-poles or
retardation.  Complementary MSP profiles borrowed from the literature
for the sake of completeness of our work were aligned by visual
inspection with the high frequency profiles.  
All MSPs have been regularly timed at various observatories in the
frequency range between 100 and 1400 MHz, which overlaps with the
frequency coverage in our study. Irregular frequency dependent timing
behaviour has not been reported so far\footnote{We note that Stairs et
al.~(1998)\nocite{sac+98} discuss an unusual frequency behaviour of the
TOAs for B1534+12 which seems however to be caused by instrumental
effects.} (see e.g.~\cite{cs84} for PSR B1937+21 as the most strongest
test).  We therefore conclude that, within the resolution and
uncertainties of our measurements, we do not observe any abnormal
frequency dependent timing behaviour over whole the radio spectrum observed
for MSPs. 

\subsection{Magnetic field structure}

This absence of any timing irregularities directly reflects on the
magnetic field topology in the radio emission region, as such could
indicate the existence of aberration effects or magnetic multi-poles
(cf.~\cite{pw92}, \cite{kxj+97}). Hence, an undisturbed, dipolar
structure of the magnetic field would be an important result for
pulsar emission physics, since the existence of non-dipolar field
components has been speculated by many authors (e.g.~\cite{cr93}) in
the past and employed to account for unusual emission phenomena.
Among these are (a) the smaller than expected emission beams of MSPs
in comparison to normal pulsars, (b) the distorted polarization
position angle curves which often exhibit flat slopes, (c) the
abnormal profile frequency development and (d) polarization properties
(cf.~Paper I and II and references therein), or also (e) "notches"
such as identified in the profile of PSR J0437$-$4715
(\cite{nm96}).

Usually, field components of higher order than dipolar existing at the
surface decay rapidly with increasing distance to the neutron star, so
that their significance may be low in the radio emission region. 
In addition, there are however two direct contributions to the toroidal 
component of the magnetic field: displacement current (as a rotating dipole 
in vacuum also has ``swept back'' field lines) and the current of the 
outsreaming plasma.  In MSPs there must be a considerable toroidal
component present virtually everywhere in the magnetosphere (its
relative strength increases as ($r/R_{LC}$) for small $r$), so that at the
light cylinder all the contributions to the magnetic field (vacuum
dipole poloidal, vacuum dipole toroidal and due to the flowing
current) are of the same order.

Given the compactness of MSP magnetospheres these effects could be
very well relevant for the observed emission properties.  For example,
the light cylinder radius, $R_{\rm LC}$, which ultimately limits the
pulsar magnetosphere, is only 70 km for PSR B1937+21.  Hence, any
non-zero emission height may be already at a significant fraction of
$R_{\rm LC}$.  Nevertheless, as we have seen, all MSP multi-frequency
profiles align perfectly between 0.4 and 5 GHz. This can be
interpreted in two ways. Either the magnetic field in the whole
emission region is purely dipolar, i.e.~neither higher order
multi-poles nor significant toroidal field components are present, or
the emission region is so compact that we would not notice any
deviation from a dipolar structure, i.e.~all radio frequencies are
virtually emitted from the same location.

\subsection{Size of the emission region}

With a typical timing accuracy of about $50\mu$s depending on
resolution and S/N, the simultaneous arrival of the multi-frequency
profiles sets a limit to the size of the emission region corresponding
to a light-travel distance of only 2.4 km.  Estimating the size of the
emission region in this manner is rather an oversimplification due to
possible non-linear propagation effects in pulsar magnetospheres
(e.g.~\cite{ba86}, \cite{bgi93}). Therefore, the size of the emission
region is also often estimated from the change in profile width
(e.g.~\cite{cor78}, \cite{pw92}, \cite{xkj+96}). This approach is
based on the observation that the component separation and profile
width of normal pulsars typically decreases strongly from low to high
frequencies (see e.g.~\cite{srw75} or \cite{pw92}).
This behaviour has been explained by the assumption
of a frequency dependent location of the emission region within the
magnetosphere, a so-called radius-to-frequency mapping (RFM, Cordes
1978).  The location is changing with frequency either in altitude
above the surface or in radial distance from the magnetic axis. (The
latter appears less likely given that profile width and component
separation change simultaneously.) With the additional assumption that
the profile wings reflect the extent of the open (dipolar) field line
region, the width can be directly converted into the height and size
of the emission region (cf.~Paper I).

An alternative explanation for the narrowing of the profile width has
been given in terms of a propagation effect involving refraction
effects and birefringence of the plasma above the polar cap
(\cite{ba86}, \cite{mck97}, \cite{gal98}, \cite{hlk98}). In this model
the emission is created at a single altitude, but not all emission can
leave the magnetosphere directly. Instead, one polarization mode first
propagates through the plasma before it escapes the magnetosphere at a
certain radius. The extent of the open field line region at this
particular, frequency dependent altitude will then be the factor which
determines the observed profile width.  Gallant (1998) gives for the
escape radius of the fast O-mode, $r_{\rm esc}$, an expression
\begin{equation}
\frac{r_{\rm esc}}{R_\ast} \approx 350 \left( \frac{B_{12}h_4 \gamma_2}{P}
\right)^{1/3} \nu_{\rm GHz}^{-2/3}
\end{equation}
where $R_\ast$ is the neutron star radius, $P$ the spin period and
$B_{12}$ the surface magnetic field.  Interestingly, unless the
multiplication factor between the primary and secondary pair plasma,
$h=10^4 h_4$, and the $\gamma$-factor of the secondaries,
$\gamma=10^2\gamma_2$, are very low, this escape radius lies outside
the light cylinder radius for most of the MSPs. If the given
approximation is correct, one may expect as a consequence to observe
only one orthogonal mode of the polarization state. This is generally
not the case. Even in the case of normal pulsars, the model implies a
separation of the orthogonal modes in pulse phase, but there is clear
evidence that the modes occur simultaneously at the same phase
(e.g.~\cite{ms98}). In Paper II this occurrence of orthogonal modes
was also demonstrated for MSPs, where we observe a discontinuity
in the swing of the position angle and a drop in the degree of linear
polarization for some sources (see also Sallmen 1998 and Stairs et al.~1999).
Besides, according to the above expression the profile
width should exhibit a frequency dependence of $\propto \nu^{-2/3}$
(modulo some possible geometrical effects).  In stark contrast, our
observations suggest a minimal frequency dependence for profile widths
of MSPs. The profile width reaches saturation either at very low radio
frequencies or does not change with frequency at all.

It is important to emphasise that both models, RFM and birefringence,
predict a change in profile width with frequency. The difference,
however, is that in the RFM model the profile width reflects the
altitude at which the emission is {\em created}, while in the 
birefringence model it
is the altitude where the emission {\em escapes} the magnetosphere.
The birefringence model has certainly the advantage that it can
explain the occurrence of orthogonal polarization modes in a natural
way. Moreover, the RFM model requires the assumption that the overall
broad-band pulsar emission mechanism exhibits also some narrow-band
features in order to produce a distinct mapping of frequency to
altitude, which seems difficult to fulfil simultaneously. Obviously,
it appears impossible to decide between these models simply on the
basis of profile width data.  Using additional timing data, we can
only set an upper limit on the size of the emission region, but the
data are in fact consistent with a single emission altitude (see also
\cite{pw92} and \cite{kxj+97}).  Blaskiewicz, Cordes \& Wasserman
(1991)\nocite{bcw91} and von Hoensbroech \& Xilouris (1997)
\nocite{hx97a} have tried to determine any dependence of the emission
height on frequency by analysing polarization data of normal pulsars
including relativistic corrections to the rotating-vector-model
(\cite{rc69a}). Although this method yields {\em negative} emission
heights for 5 out of 36 sources (Blaskiewicz et al.~1991, von
Hoensbroech \& Xilouris 1997), and although most of the results are,
within the (often large) uncertainties, consistent with a constant
emission altitude, there is an interesting trend visible in the
data. According to the interpretation of the Blaskiewicz et al.~model,
this trend {\em suggests} a weak but non-zero RFM. However, any {\em
evidence} has yet to be obtained and a single emission altitude is
still consistent with the data. The pulsar width data of MSPs
presented here at least indicate a very weak or even non-existing RFM
for these sources, but this may not be surprising given the
compactness of their magnetospheres --- even if RFM is present in
normal pulsars.

\subsection{De-polarization and profile width}
 
As mentioned before, the occurrence of orthogonal polarization modes is
a natural consequence of the birefringence model. An overlap of
orthogonal modes leads generally to a de-polarization of the average
pulse profile. The emission of normal pulsars is in fact well known to
exhibit a strong de-polarization with frequency (e.g.~\cite{mth75}) and
observations at very high radio frequencies (i.e.~32 GHz) suggest that
all pulsars become de-polarized at sufficiently high frequencies
(\cite{xkj+96}). Among other models (see e.g.~\cite{xkjw94} for a
summary), it was therefore suggested that an increased occurrence of
orthogonal modes at high frequencies may be responsible for the observed
de-polarization (\cite{scr+84}). McKinnon (1997)\nocite{mck97} finally
linked the decreasing profile width and the increasing de-polarization
as a function of frequency to the same birefringence model. 

Our polarimetric profiles of MSPs discussed in this work suggest that
we observe an essentially complete de-polarization of MSPs already at
frequencies around 5 GHz -- at least for the limited sample with
polarization data up to this frequency. At the same time, even though
very little is known about single pulse polarimetry of MSPs (see
e.g.~Sallmen 1998), it is clear that orthogonal polarization modes are
present in the emission of MSPs (e.g. \cite{ts90}, Paper II, Sallmen
1998, Stairs et al.~1999). However, the profiles of MSPs show a
strikingly constant width across the observed frequencies.  In other
words, the de-polarization of MSP emission appears to be de-coupled
from the effect of profile narrowing.  Since MSPs and normal pulsars
appear to share the same emission mechanism (cf.~Sect.~\ref{intro}),
we are therefore led to the conclusion that these phenomena are
unrelated for normal pulsars as well, at least to zero order.

\subsection{Flux density spectra}

The apparent simplicity of the observed flux density spectra may also
point to a small size of the emission region, perhaps embedded in a
dipolar structure of the magnetic field. One could expect that a
different field geometry or at least a significant change in emission
height also alters the efficiency of the emission process, e.g.~due to
a change in magnetic field strength, curvature radius or plasma
densities (see e.g.~Kuzmin et al.~1986\nocite{kmi+86}). It is
therefore interesting that we very often observe a steepening in the
flux density spectra of normal pulsars but apparently not for MSPs. In
fact, at least 55\% of all 45 normal pulsars studied by Malofeev et
al.~(1994)\nocite{mgj+94} exhibit a break in their power law spectra
with a mean break frequency of $2.2\pm0.4$ GHz.  From the twelve MSPs
with flux measurements up to 5 GHz, only PSR J0437$-$4715 shows a
clear steepening in its spectrum above 1 GHz, while PSRs J1713+0747
and J2145$-$0750 are likely cases.  Assuming that the fraction of
pulsars with double power law spectra is the same for MSPs and normal
pulsars, we could expect about six MSPs in our sample to exhibit a
steepening spectrum. It is important to note that the statistical
significance for this difference being real may be still low as
indicated by a $\chi^2$-test which yields only a 60\% probability that
both samples are different. However, treating those spectra which are
consistent with a straight power law as such with a break frequency at
zero Hz, we can compare the spectral break frequency distribution of
normal pulsars and MSPs. A Kolmogorov-Smirnov test then yields the
result, that both samples are drawn from the same mother distribution
with a probability of only 2.2\%. It is also interesting to compare
the mean spectral index derived for MSPs detected at 4.9 GHz to those
of 108 normal pulsars between 1.4 and 4.9 GHz which were measured with
the same observing system by Kijak et al.~(1998)\nocite{kkwj98}.  The
MSP spectral index of $-1.6\pm0.1$ is somewhat flatter than that of
normal pulsars for which the mean spectral index is $-1.9\pm0.2$ above
1.4 GHz, although we did not apply any spectral criterion when
selecting our sources.

An interesting spectral behaviour was already indicated by the early
work of Foster et al.~(1991) who studied the spectra of PSRs
B1620$-$26, B1821$-$24, B1855+09 and B1937+21. Their analysis
including also low frequency measurements (cf.~Fig.~\ref{specfig})
revealed simple straight power laws from very low frequencies up to a
few GHz, without indications of any low-frequency turn-over. In fact,
according to Erickson \& Mahoney (1985)\nocite{em85}, there is not any
indication of a maximum or turn-over in the spectrum of PSR B1937+21
above 10 MHz. The flux measurement at 102 MHz for PSR J2145$-$0750 by
Kuzmin \& Losovsky (1996)\nocite{kl96} shown in Fig.~\ref{specfig}
establishes this trend for yet another source. However, more low frequency
flux measurements are of utmost importance to investigate this
question further.

\subsection{Profile evolution -- Cones \& Cores}

We have reviewed in Paper I that the large number of pre- and
post-cursors and interpulses in MSP profiles is a very significant
difference to the emission properties of normal pulsars. In Paper II
we revealed similarly differences in the profile evolution of MSPs and
normal pulsars. These findings are confirmed in this work covering a
larger frequency range.

The profile evolution reflects the spectral behaviour of individual
profile components. In the case of normal pulsars, we observe a very
clear dependence of the component spectral index on the distance to
the magnetic axis (\cite{ran83a}, \cite{lm88} or \cite{kwj+94}).
Although central components in normal pulsar profiles can be often
associated with so-called ``core'' components (\cite{ran83a};
\cite{lm88}), the usually steeper spectral index can also be found for
some central components which are certainly so-called ``conal''
components (\cite{lm88}; \cite{kwj+94}).  Although we can distinguish
core and conal components also by their polarisation properties
(see \cite{ran83a}) even in the case of MSPs (see Paper II), one should
obviously be reluctant to simply identify such obvious differences in
{\em emission properties} directly with different {\em emission
processes}.  Other explanations for the steeper spectrum like a
geometrical origin are equally likely (\cite{kwj+94}, \cite{sie97}).
Whatever is responsible for the observed patterns in emission
properties of normal pulsars, it is clear from the profile
developments discussed in Paper II and in this work that the canonical
picture of strong central components at low and strong outer
components at high frequencies does not simply apply to MSPs. Although
some examples for this trend can be even found among the sample of
MSPs (e.g.~PSRs J0751+1807 and B1913+16), for many sources this is not
true and leads either to no profile development at all (e.g.~PSRs
J0621$-$1002 and J1744$-$1134) or to one which was classified as
``abnormal'' (relative to our expectations originating from studies of
normal pulsars) in Paper II (e.g.~PSRs J0437$-$4715 or J2145$-$0750).

In some cases where we consider the profile development as abnormal,
we might have been mislead to this classification by the possibility
that we observe a profile type which would be classified as ``partial
cone'', following the terms of Lyne \& Manchester (1988; see their
Table 4). In such a case, like for the normal pulsar PSR B0355+54,
only the leading (or trailing) part of the full emission cone is
active, so that a core component is located at the
trailing (or leading) outer edge of the visible profile.  As a
consequence, this apparent outer (but truly central) component would
generally show a steeper flux density spectrum. Pulsar J0613$-$0200 is
almost certainly such a case, since the polarization profiles suggest
that the trailing component is a core, which also shows by far the
steepest spectrum (see Sallmen 1998 and Stairs et al.~1999).
Therefore, the study of the profile development of MSPs shows that
unfilled emission beams are certainly present. This fact may help to
solve the problems of the undersized emission beams discussed in
detail in Paper I and the overall flat position angles which were
first noted in Paper II (see Sallmen 1998).  However, we are still
left with the fact, that typical patterns in profile development are
hardly observed for MSPs.

The lack of clear trends visible in the profile evolution of MSPs
(with frequency) is a good indication that either the emission
structures are not as regular as for normal pulsars (e.g.~due to
unfilled emission beams), or that the formation of recognisable
patterns is only possible if the emission region provides enough
space.  Again, the compactness of the magnetosphere might thus prevent
the formation of profile components with distinct emission properties.
The presence of disturbing effects like swept-back magnetic fields or
magnetic multi-poles is still possible but less likely given the good
alignment of the pulse profiles across the frequencies.  Under the
presumption that the emission mechanism of MSPs and normal pulsars is
the same, as supported by the results of Paper I \& II and also Jenet
et al.~(1998), this study of MSPs suggests that the different
characteristics of core and conal components found for normal pulsars
are in fact due to differences in emission heights or size of the
emission region or due to geometrical or propagation effects in the
magnetosphere, but not due to a fundamentally different emission
process. This conclusion is in agreement with that of Lyne \&
Manchester (1988) which recently found new support by the work of
Manchester et al.~(1998)\nocite{mhq98}, who presented a large number
of polarization profiles of normal pulsars, where ``typical'' core
polarization properties can be sometimes observed in cone components.

\section{Conclusions}

Considering the frequency independence of the profile width and the
observed de-polarization of the radiation, MSP emission properties
tend to resemble those of normal pulsars only shifted towards higher
frequencies.  This trend is also consistent with the early termination
of the profile development in frequency. All these characteristics may
be understood in the context of a very compact magnetosphere existing
for MSPs which possibly contains only a extremely small emission
region.  Most likely, these regions are too small to allow for 
timing irregularities or a significant change in profile shape or 
width or perhaps flux density spectrum.

While the profile width and component separation appear remarkably
constant for all frequencies, we cannot ultimately decide in favour or
against a RFM, which would be certainly weak for MSPs. However,
polarization observations for a few sources up to 5 GHz suggest that
the decrease in profile width and de-polarization of pulsar emission
are not directly related.

We find weak indications that the spectra of MSPs may not show the
spectral break as often observed for normal pulsars around a few GHz. At
the same time, we cannot observe a clear dependence of spectral index
of individual pulse components on distance to the profile midpoint
(and presumably magnetic axis). Nevertheless, features like partial
cones as known from normal pulsars seem also to be present for some
MSPs. A combination of all available MSP data with the previously
shown fact that the radiation of MSPs and normal pulsars is obviously
created by the same emission mechanism, leads us to conclude
that there is a no direct evidence for a fundamentally different
emission process of so-called core and conal components.

Finally, this study demonstrates that for a number of MSPs regular
observations at frequencies above 2 GHz are easily possible with a
wide bandwidth system. Besides studying emission properties, such
measurements can provide a useful tool for high precision timing by a
determination of disturbing effects due to the interstellar weather
(\cite{bw99}).

\acknowledgements{During this work we made {\em extensive} use of the
European Pulsar Network data archive. We are thus indebted to all
authors who generously contributed their measurements to the database.
MK acknowledges the receipt of the Otto-Hahn Prize, during whose tenure
this paper was written, and the warm hospitality of the Astronomy
Department at UC Berkeley. The authors also benefited from stimulating
discussions with J.~Arons, J.~Hibschman, A.~Melatos, S.~Sallmen,
A.~Somer, and A.~Spitkovsky. Arecibo Observatory is operated by Cornell
University under cooperative agreement with the National Science Foundation.}


\clearpage

\figcaption{\label{specfig} 
Flux density spectra for 20 millisecond pulsars discussed
in this work. For references see text.}

\figcaption{\label{fig0218} Profiles of PSR J0218+4232 at different
frequencies.  The 410-MHz and 610-MHz profiles were taken from Stairs
et al.~(1999). 
Profiles are aligned by eye since timing
information was not available for the low frequency data.}

\figcaption{\label{fig0437} Profile of PSR J0437$-$4715 at different
frequencies.  The alignment of the profiles, which were provided to
the EPN-data archive by Manchester \& Johnston (1995) and Bell et
al.~(1995), was made by eye following the measurements by Navarro et
al.~(1997).}

\figcaption{\label{fig0621} Profiles of PSR J0621+1002 at different
frequencies.  The 410-MHz profile was taken from Stairs et al.~(1999),
the 575-MHz profile from Camilo et al.~(1996).  Profiles are aligned
by eye since timing information was not available for the low
frequency data.}

\figcaption{\label{fig1012} Profiles of
PSR J1012+5307 at different frequencies. The 575-MHz profile was 
taken from Sallmen (1998). Profiles
are aligned by a timing model, confirming the weak detection at 4.9 GHz.}

\figcaption{\label{fig1022} Profile PSR J1022+1001 at different
frequencies.  The 370-MHz profile was taken from by Sayer et
al.~(1997), those at 410 and 610 MHz from Stairs et al.~(1999). Apart
from these profiles the data are aligned by a timing model.}

\figcaption{\label{fig1024} Profiles of PSR J1024$-$0709 at different
frequencies.  The 436-MHz profile was taken from the EPN-database and
was provided by Bailes et al.~(1997). Apart from this profile the data
are aligned by a timing model.}

\figcaption{\label{fig1518} Profiles of PSR J1518+4904 at different
frequencies.  The profiles at 370, 575 and 800 MHz were taken from
Nice et al.~(1996), the profile at 610 MHz from Stairs et al.~(1999).
Apart from these profiles the data are aligned by a timing model.}

\figcaption{\label{fig1643} Profiles of PSR J1643$-$1224 at different
frequencies. The profile at 610 MHz was taken from Stairs et al.~(1999).
Apart from this profile the data are aligned by a timing model.}

\figcaption{\label{fig1713} Profiles of PSR J1713$+$0747 at different
frequencies.  The 430-MHz and 2305-MHz profiles were taken from Foster
et al.~(1993), the 610-MHz profile from Stairs et al.~(1999).  Apart
from these profiles all data are aligned by a timing model.}

\figcaption{\label{fig1744} Profiles PSR J1744$-$1134 at different
frequencies.  The 410-MHz profile was taken from Bailes et al.~(1997),
the 610-MHz profile from Stairs et al.~(1999).  Apart from these two
profiles all data are aligned by a timing model.}

\figcaption{\label{fig1855} Comparison of the profiles of B1855+09
at 1.41 and 4.85 GHz. The latter was taken from Kijak et al.~(1997).
Both profiles were aligned by a timing model.}

\figcaption{\label{fig1937} Profiles of PSR B1937+21 at different
frequencies.  All profiles are aligned by a timing model.}

\figcaption{\label{fig2145} Profiles of PSR J2145$-$0750 at different
frequencies.  The profile at 102 MHz was provided by Kuzmin \&
Losovskii (1996), the profiles at 410 and 610 MHz by Stairs et
al.~(1999).  The 4850-MHz profile was taken from Kijak et al.~(1997).
All profiles above 610 MHz are aligned by a timing model.}

\figcaption{\label{singleprof}
Pulse profiles of PSRs J0751+1807
and J1640+2224 at 2.695 GHz.} 

\figcaption{\label{polfig} Degree of polarization for the linearly
polarized intensity. 
Data were compiled from Paper II, Manchester \& Johnston
(1995), Navarro et al.~(1997), Sallmen (1998), Stairs et al.~(1999),
Thorsett \& Stinebring (1990) and this work.}

\figcaption{\label{figsepwidth} Profile width measured at a 50\% intensity
level (filled symbols) and component separation (open symbols) for the
sources discussed in this work. Circles mark measurements where coherently
de-dispersed profiles were used, triangles mark values based on incoherently 
de-dispersed data. References for the used profiles are given in the text,
figure captions and Paper I and II.}

\figcaption{\label{figsepwidthhist} Comparison of index of profile
narrowing (unfilled area) and component separation change (shaded area)
for normal and millisecond pulsars. The values for normal pulsars were
taken from Thorsett (1991) and Xilouris et al.~(1996).}

\clearpage

\include{tab1}

\clearpage

\include{tab2}

\clearpage

\include{tab3}

\clearpage

\include{tab4}

\clearpage

\include{tab5}

\clearpage

\include{tab6}

\clearpage

\plotone{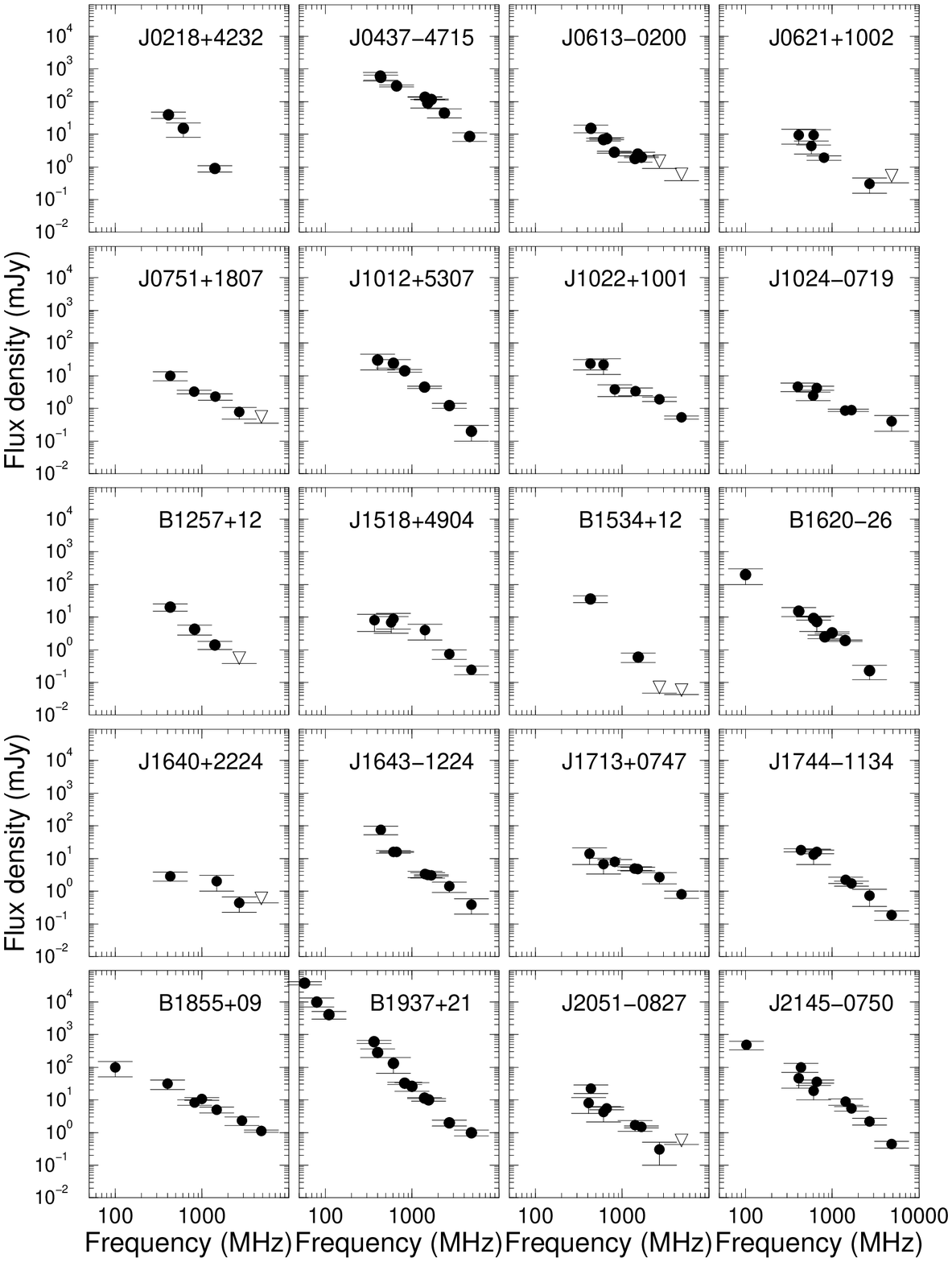}

\clearpage

\plotone{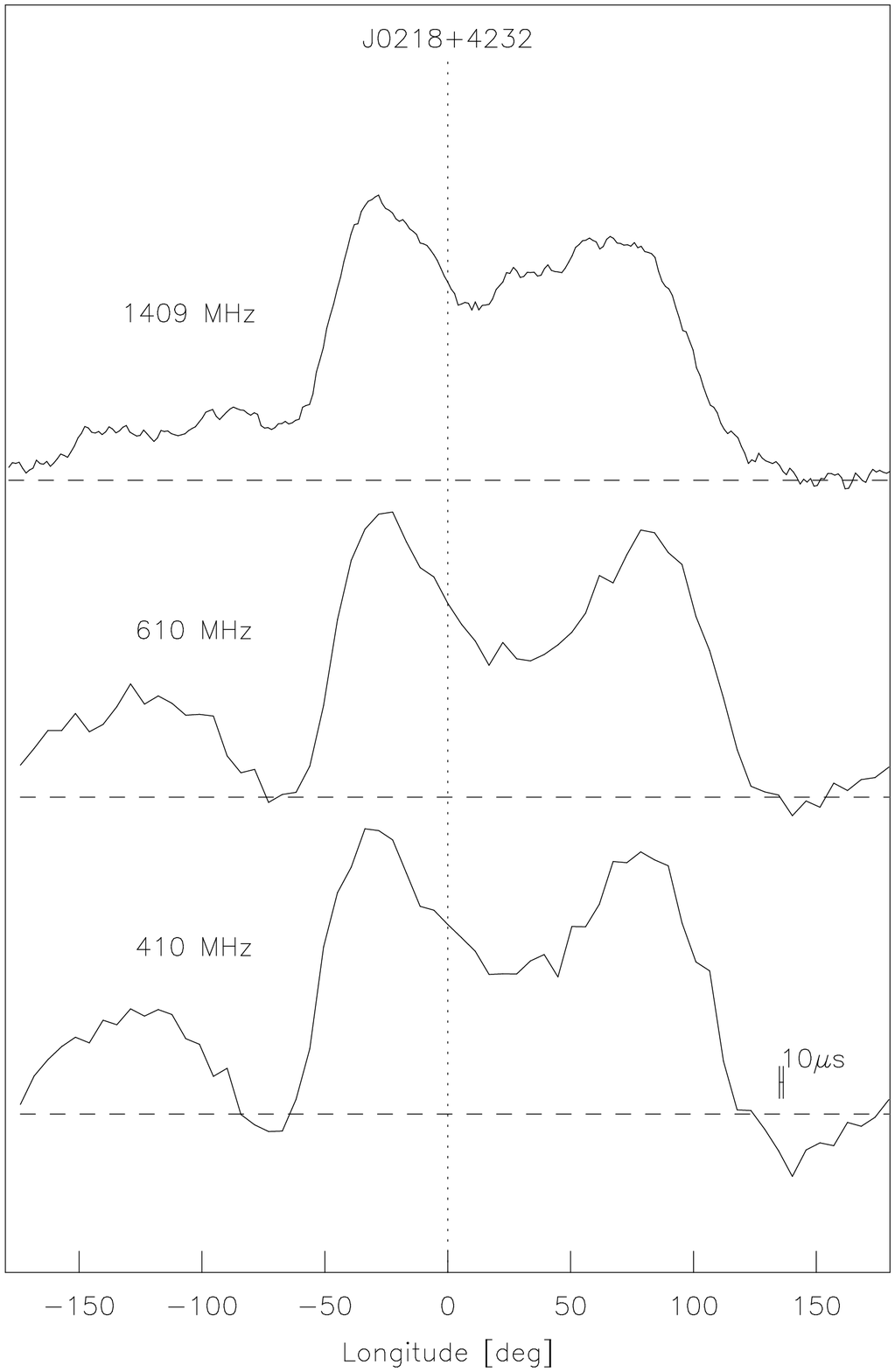}

\clearpage

\plotone{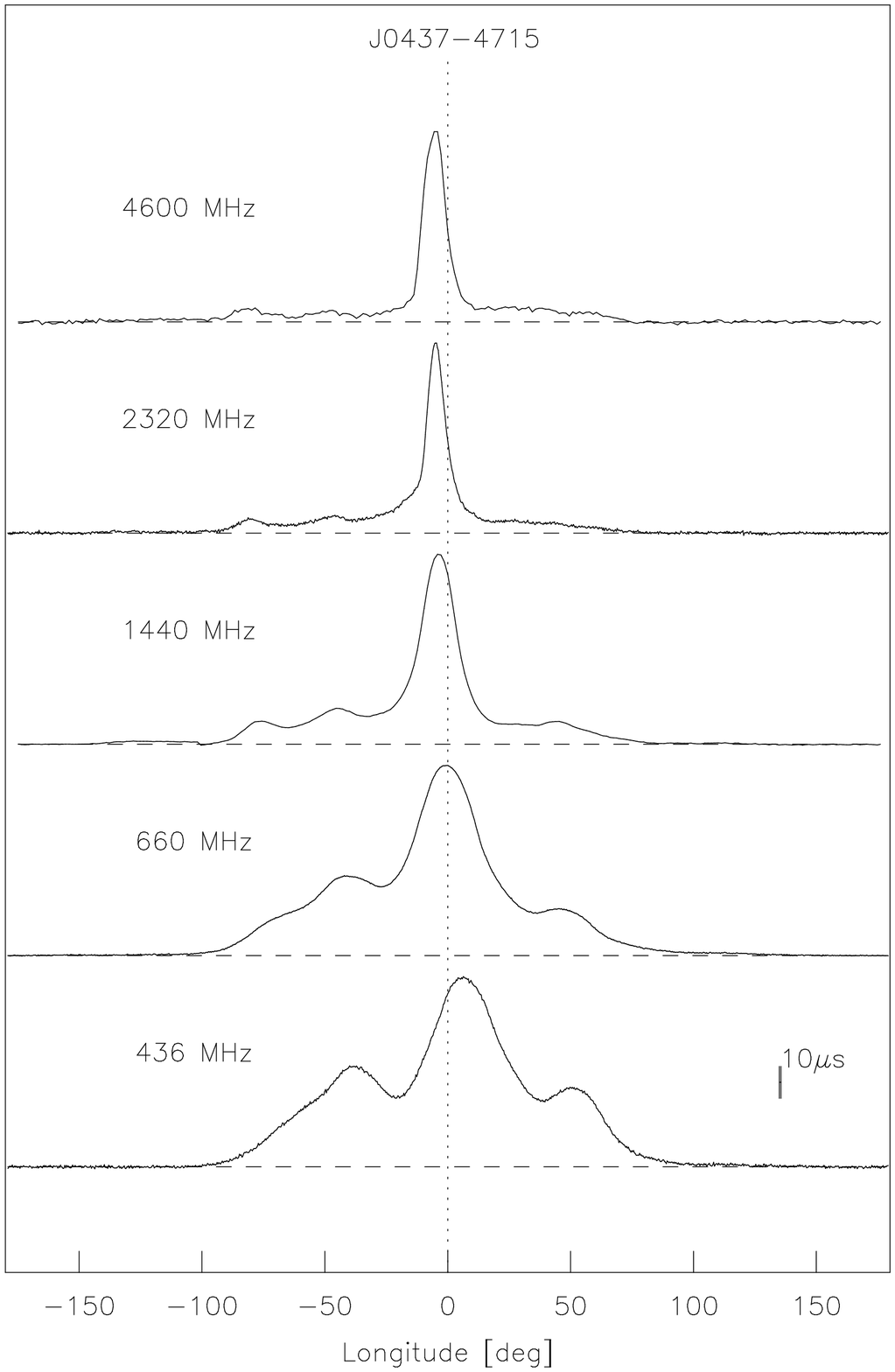}

\clearpage

\plotone{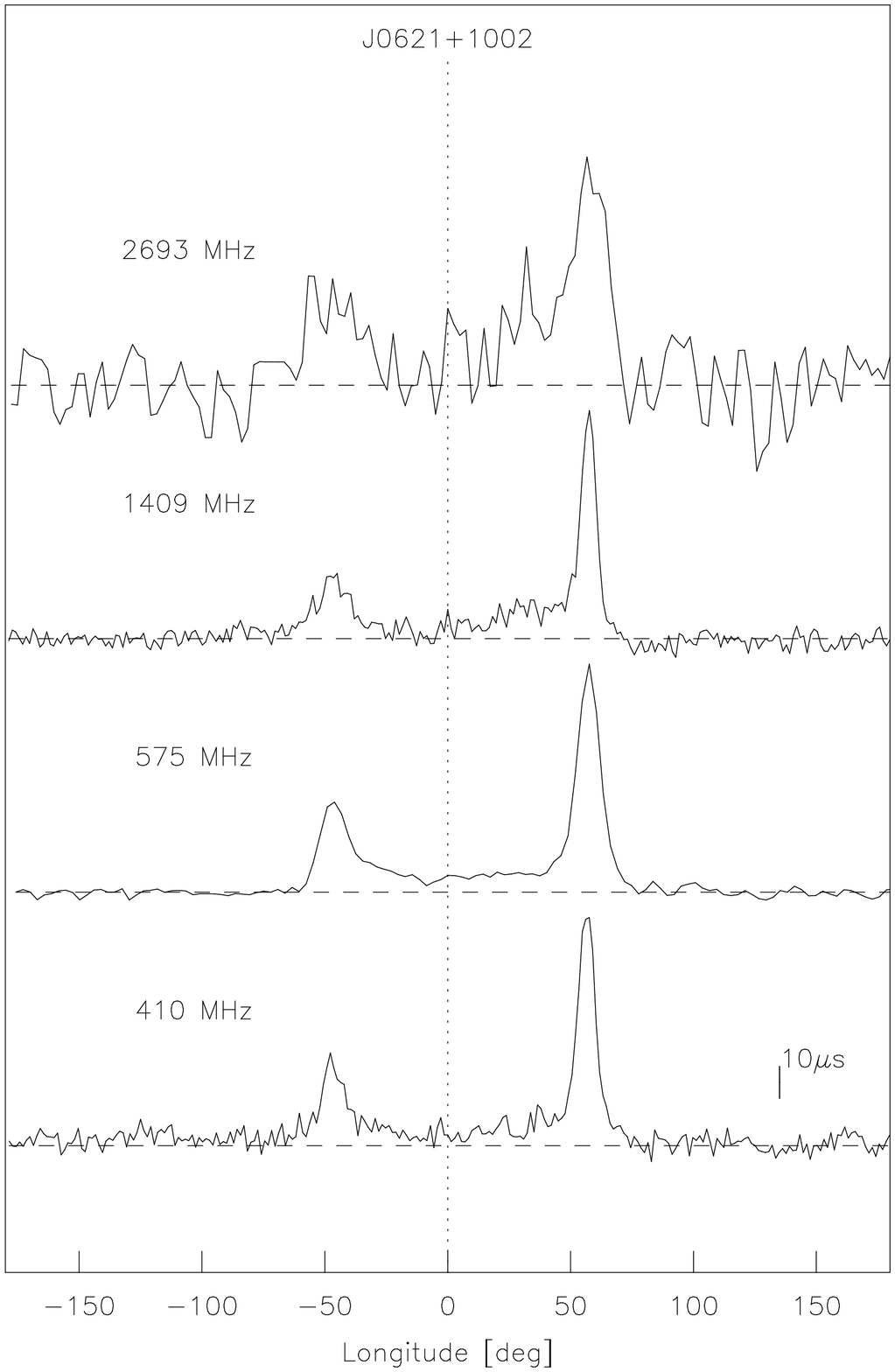}

\clearpage

\plotone{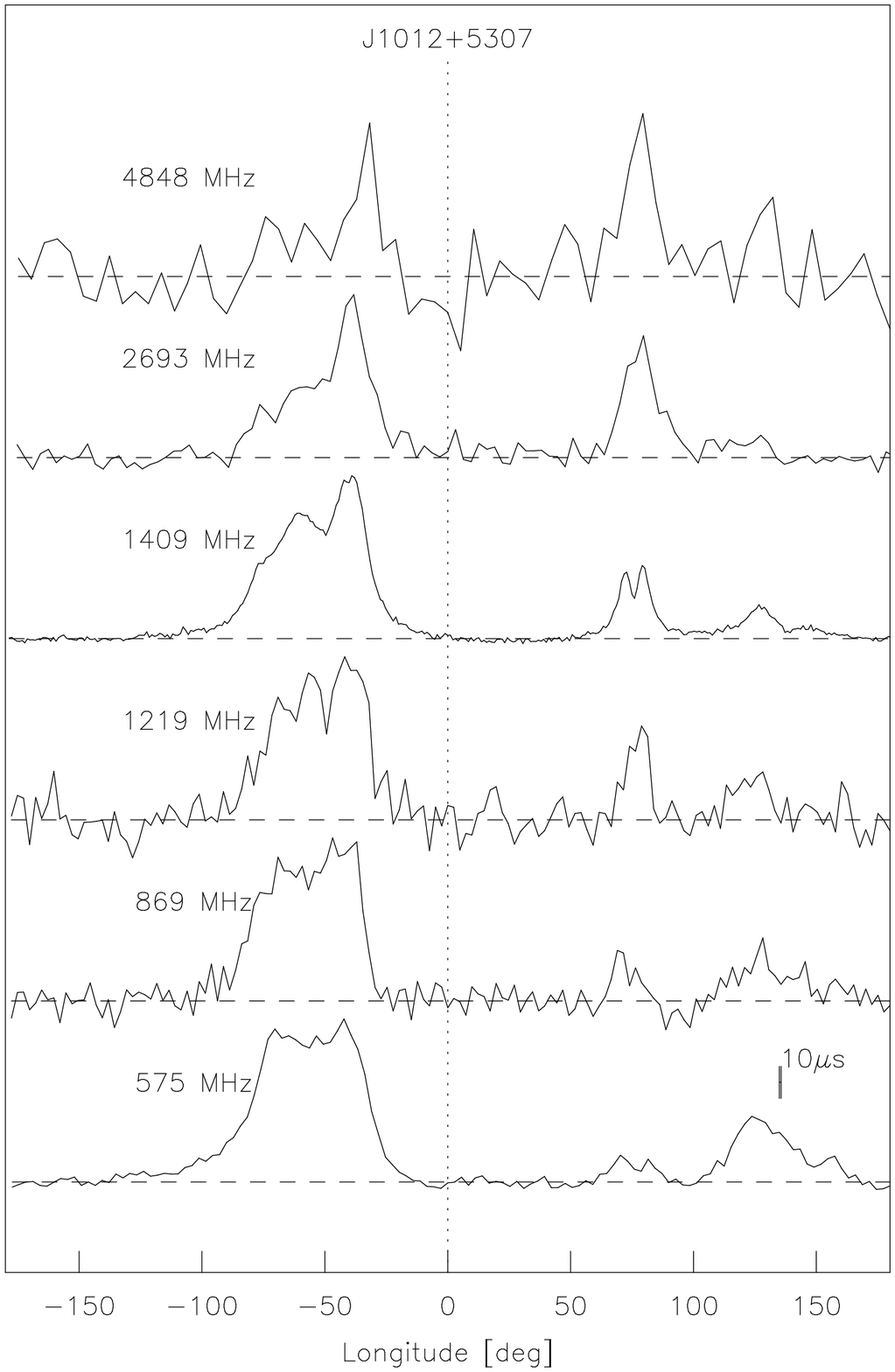}

\clearpage

\plotone{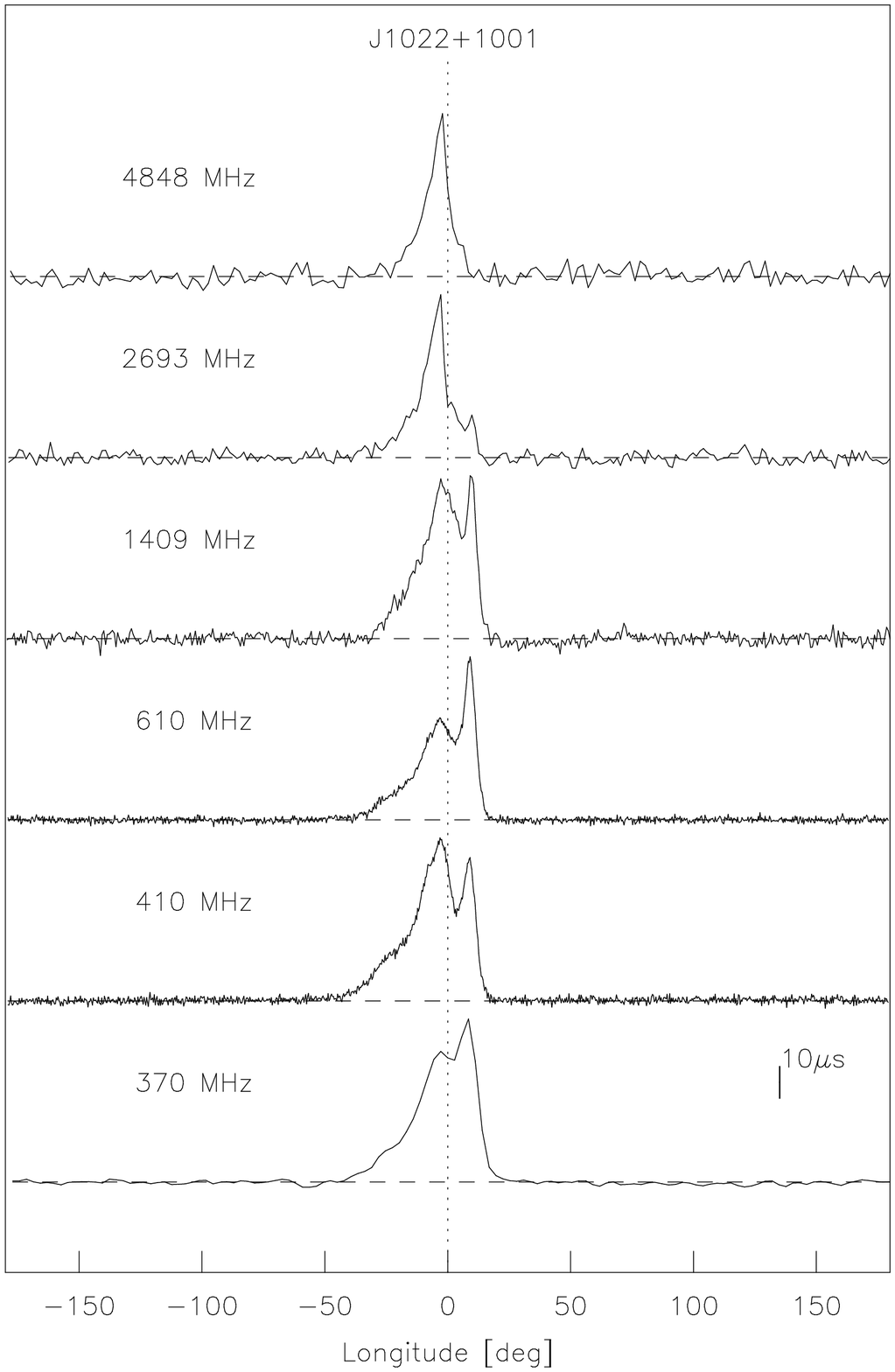}

\clearpage

\plotone{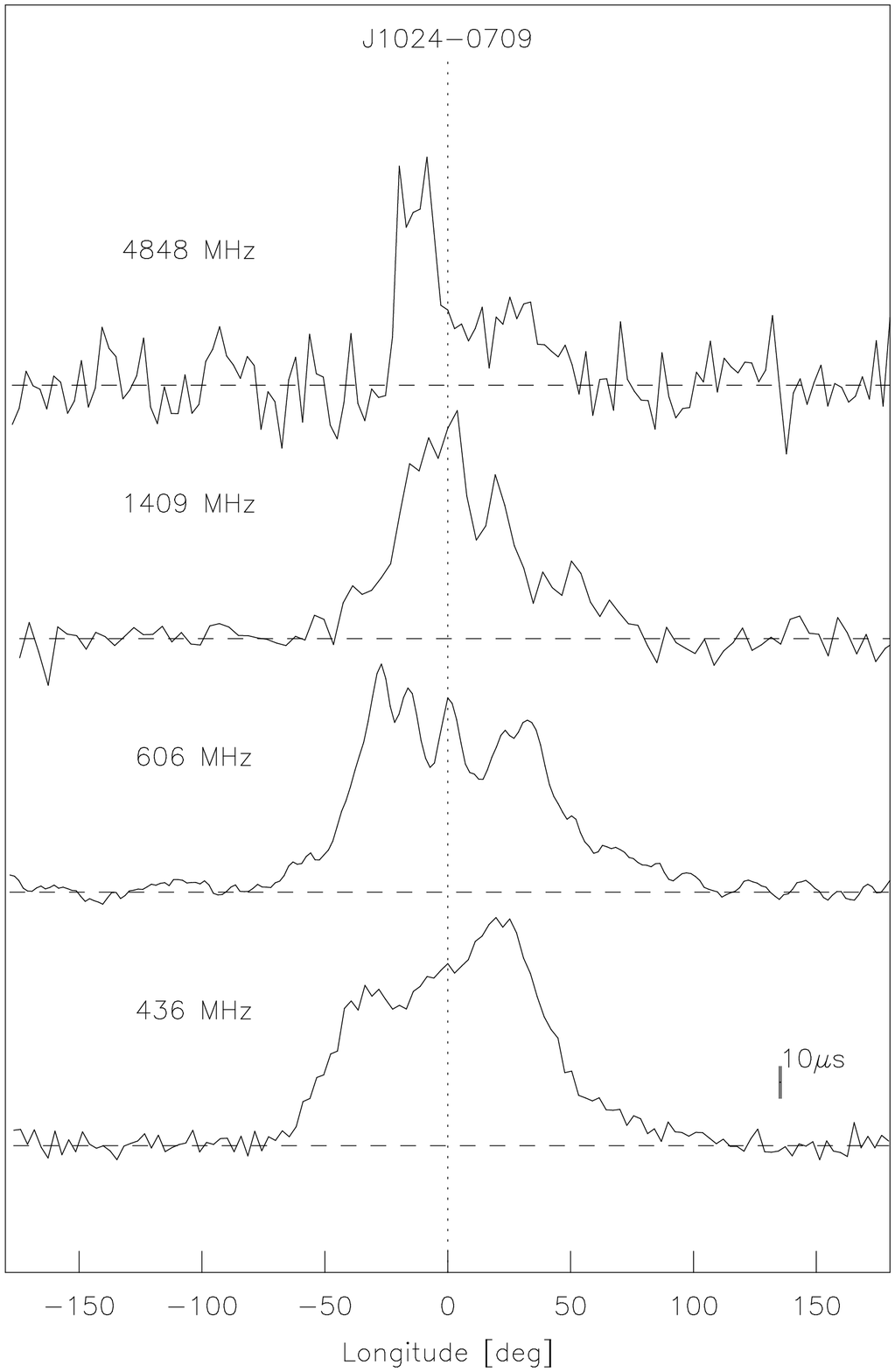}

\clearpage

\plotone{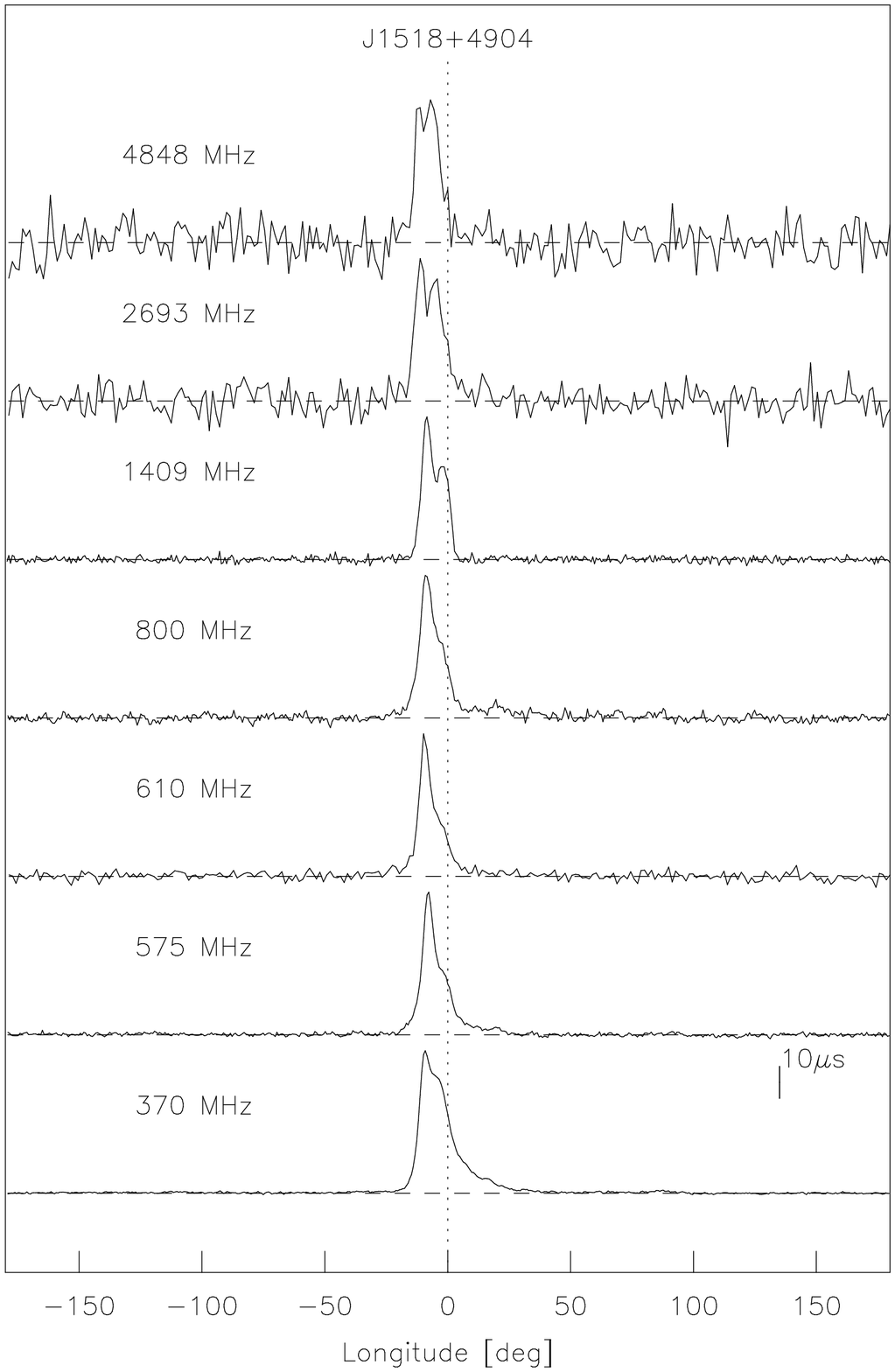}

\clearpage

\plotone{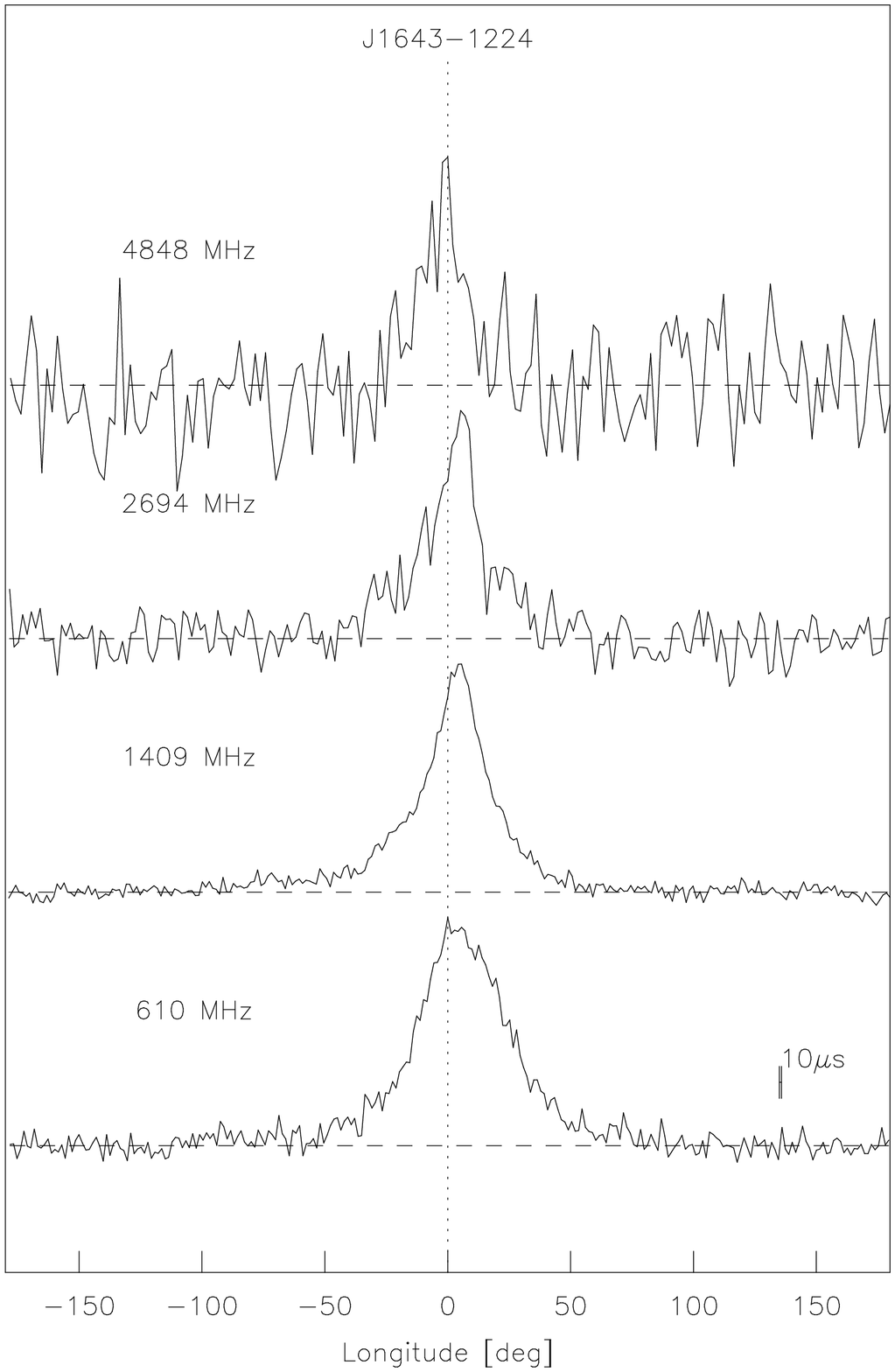}

\clearpage

\plotone{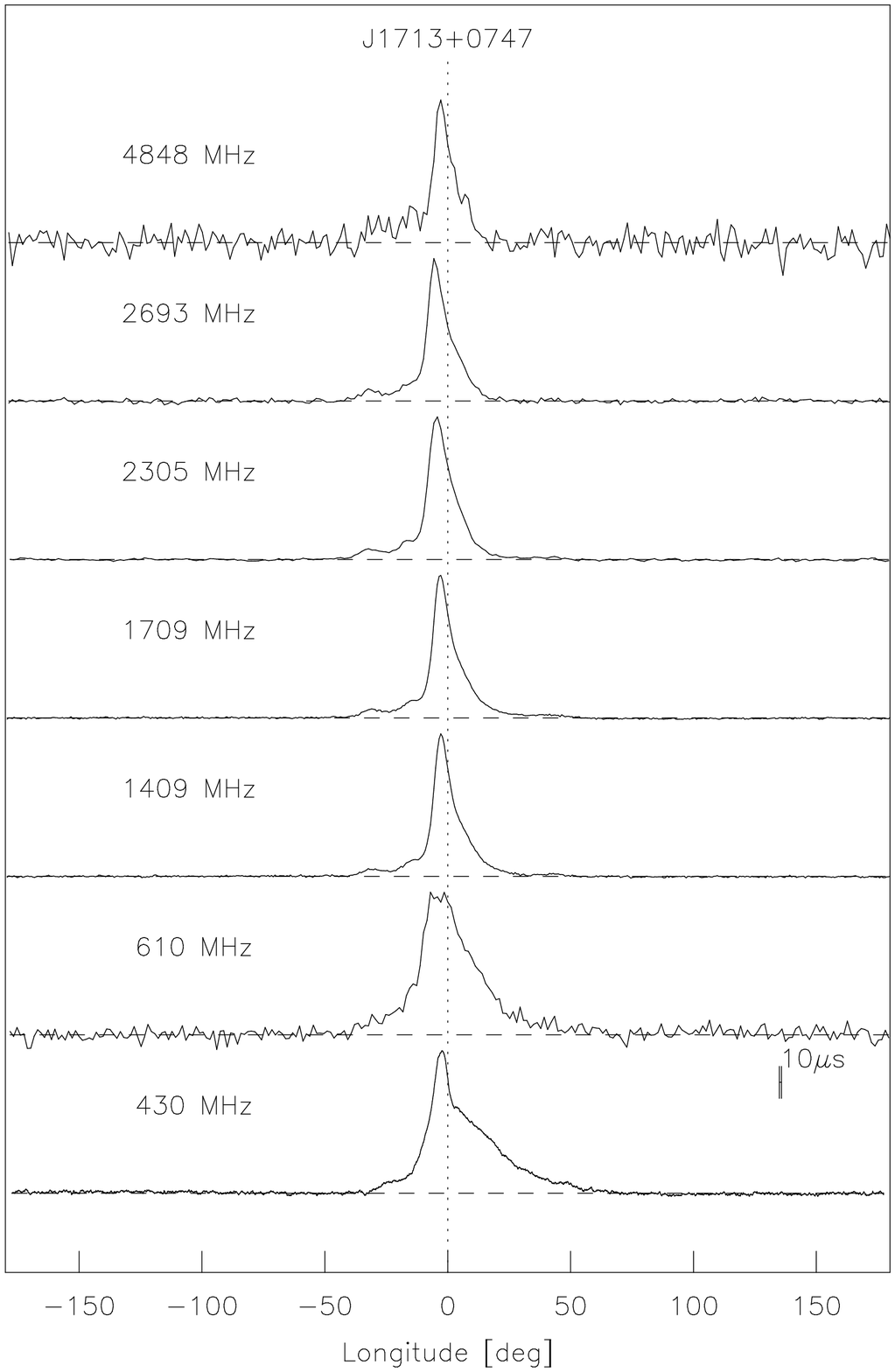}

\clearpage

\plotone{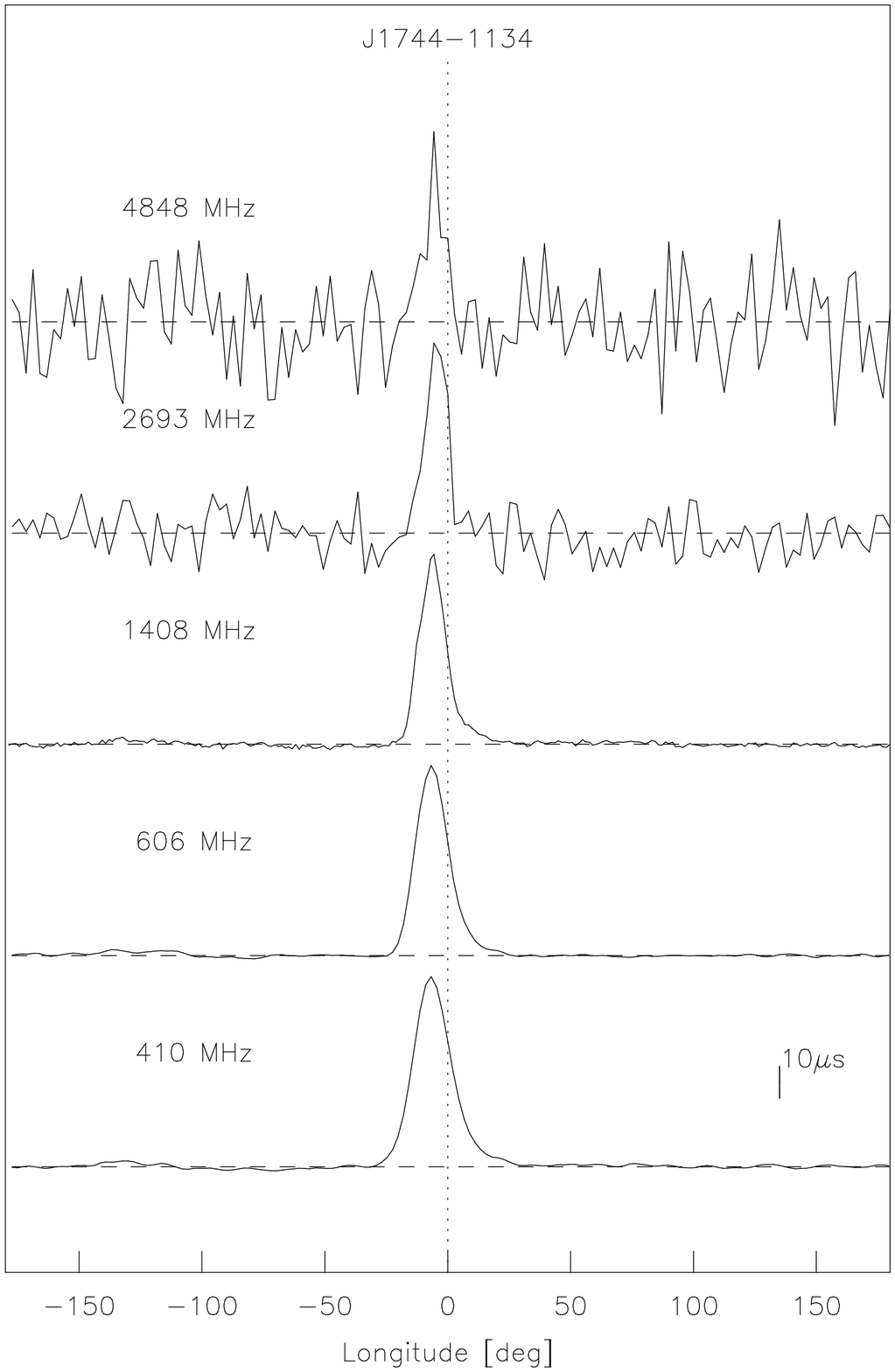}

\clearpage

\plotone{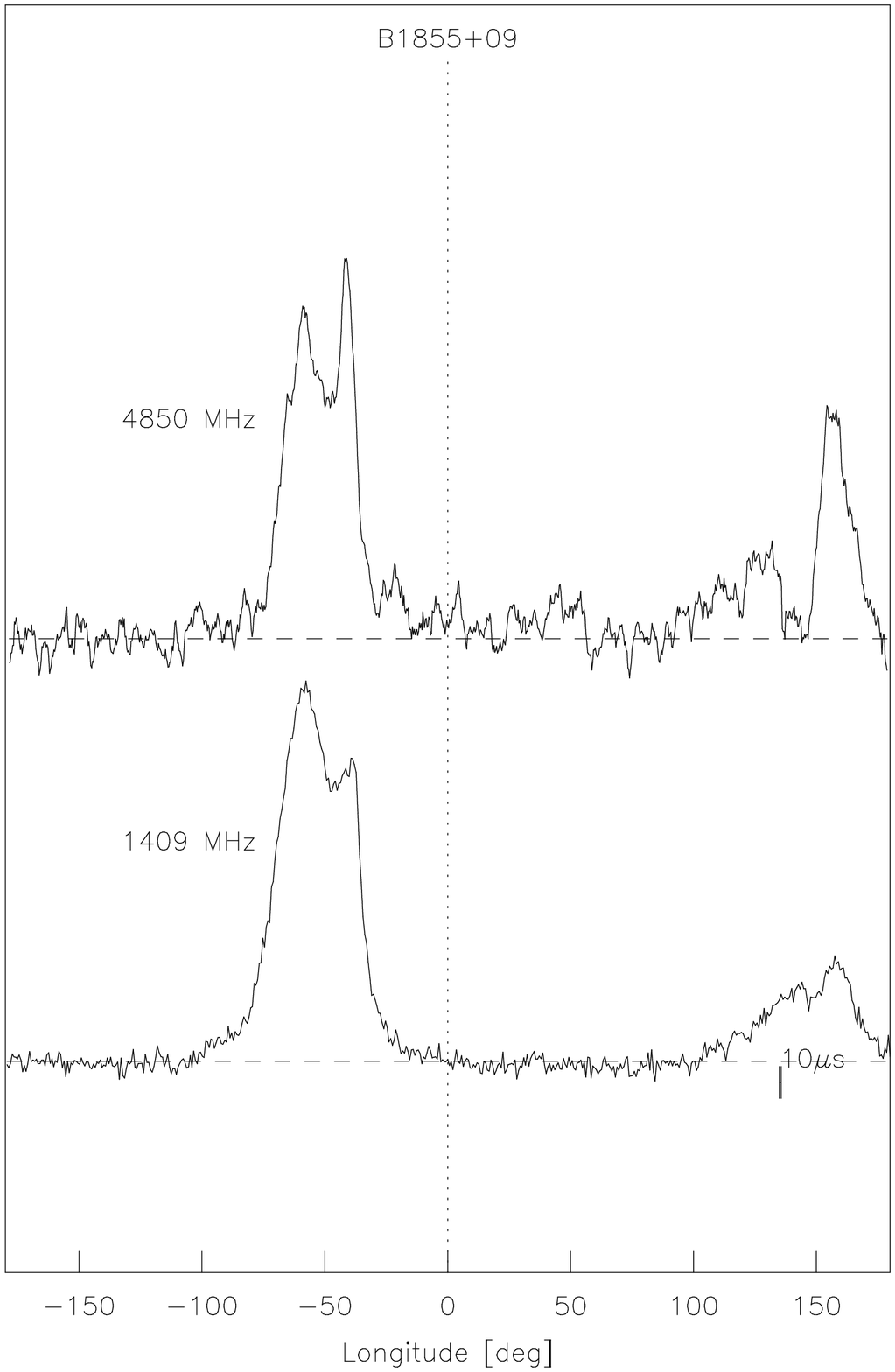}

\clearpage

\plotone{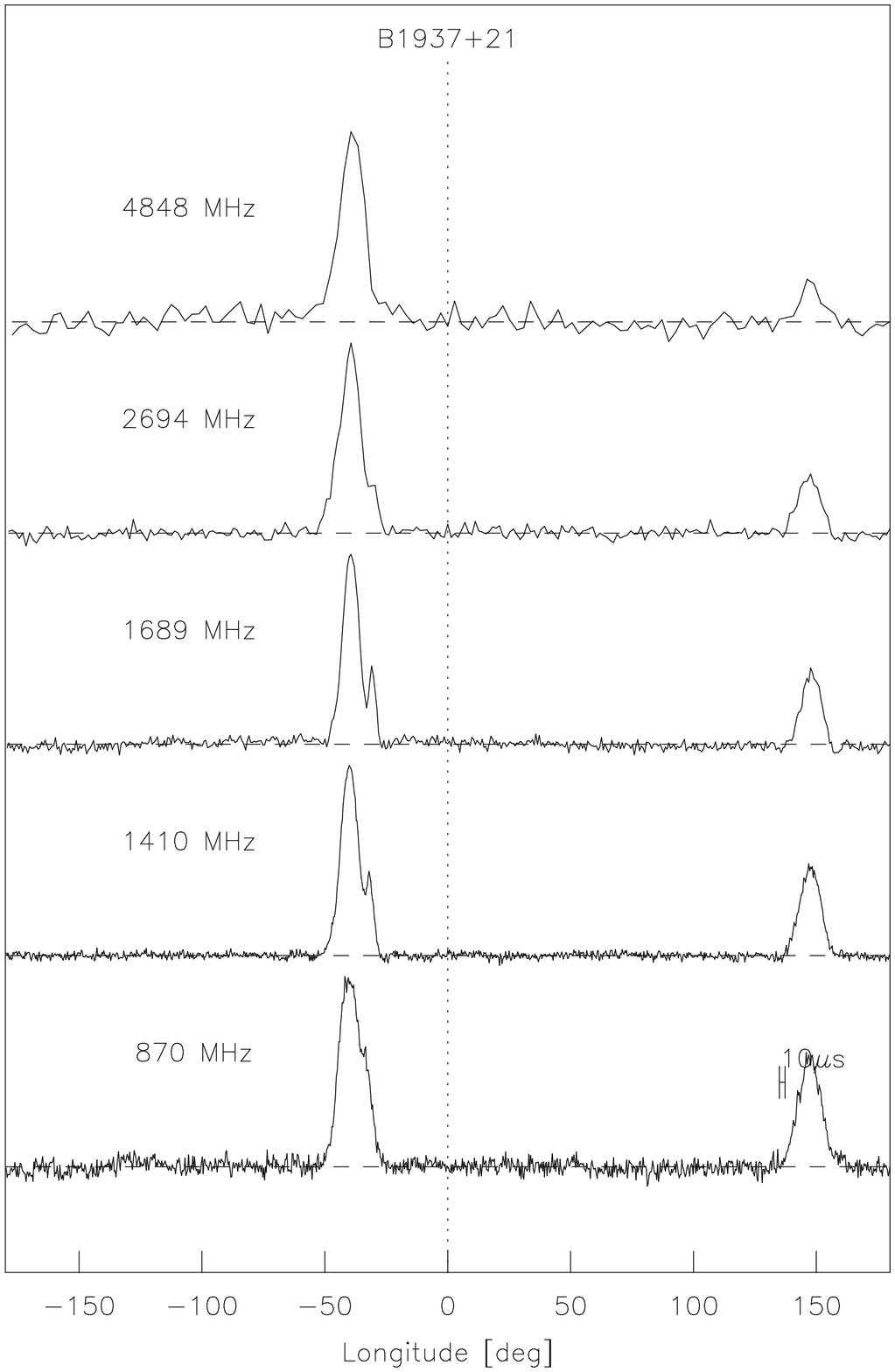}

\clearpage

\plotone{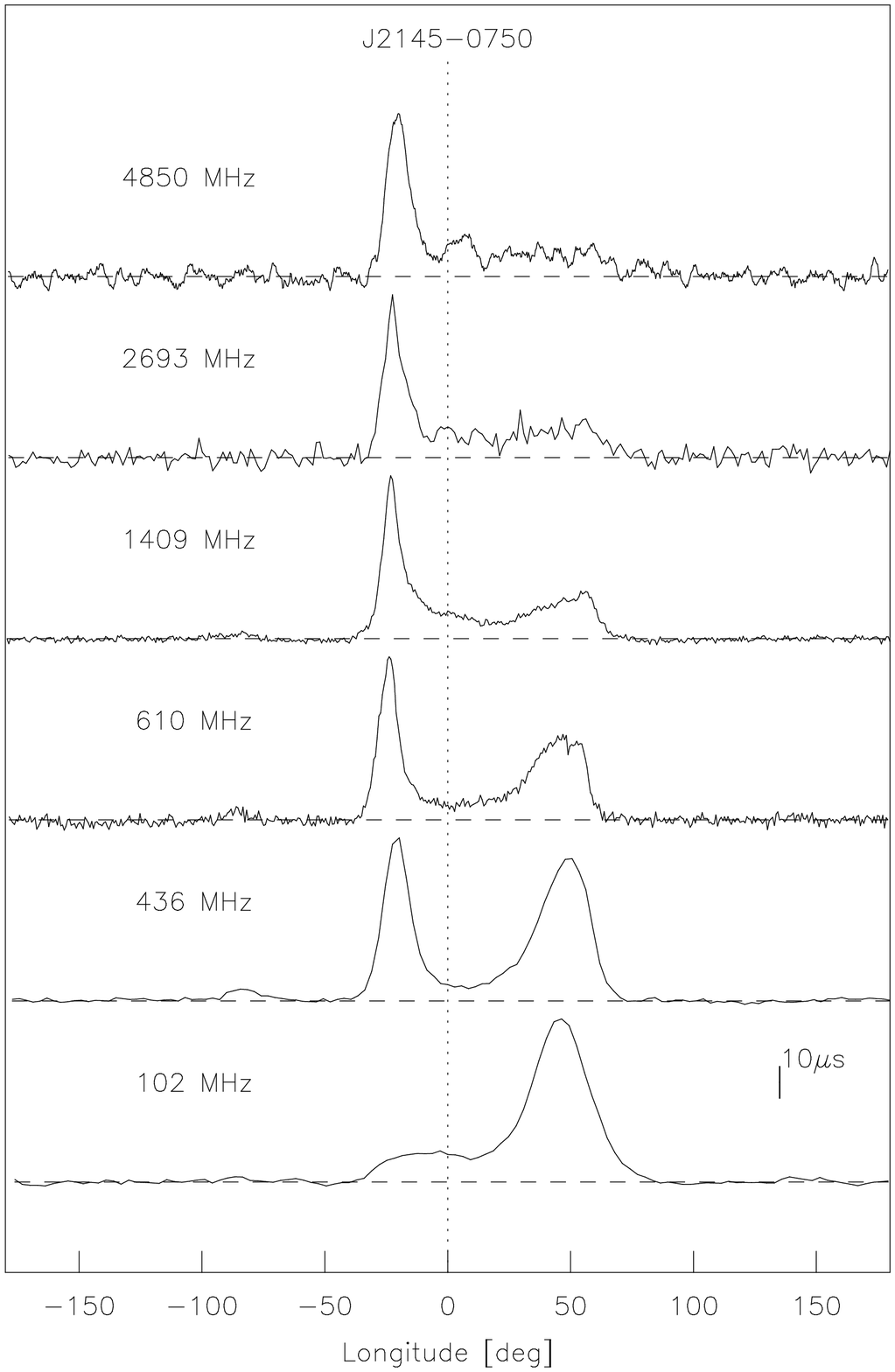}

\clearpage

\plotone{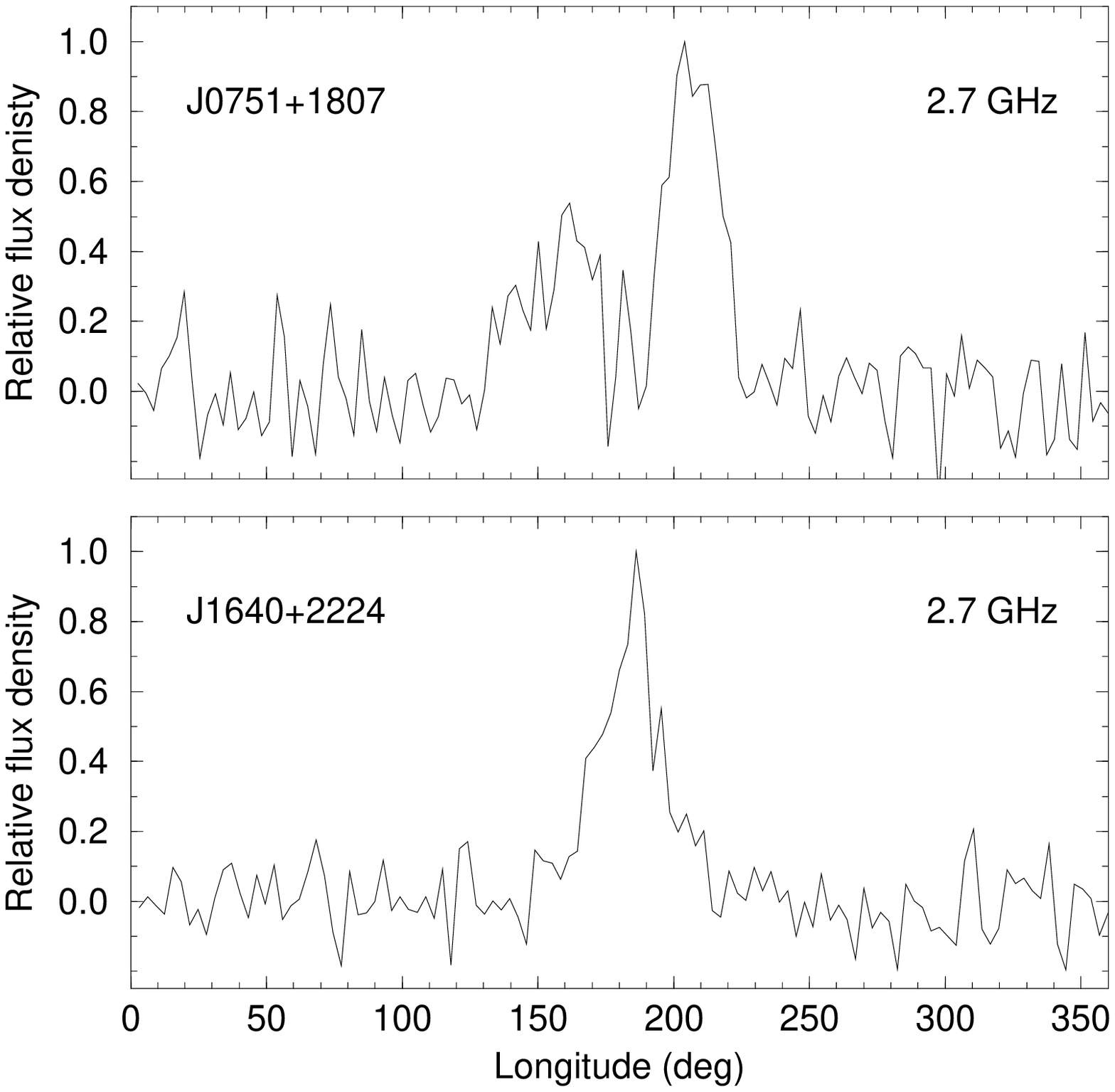}

\clearpage

\plotone{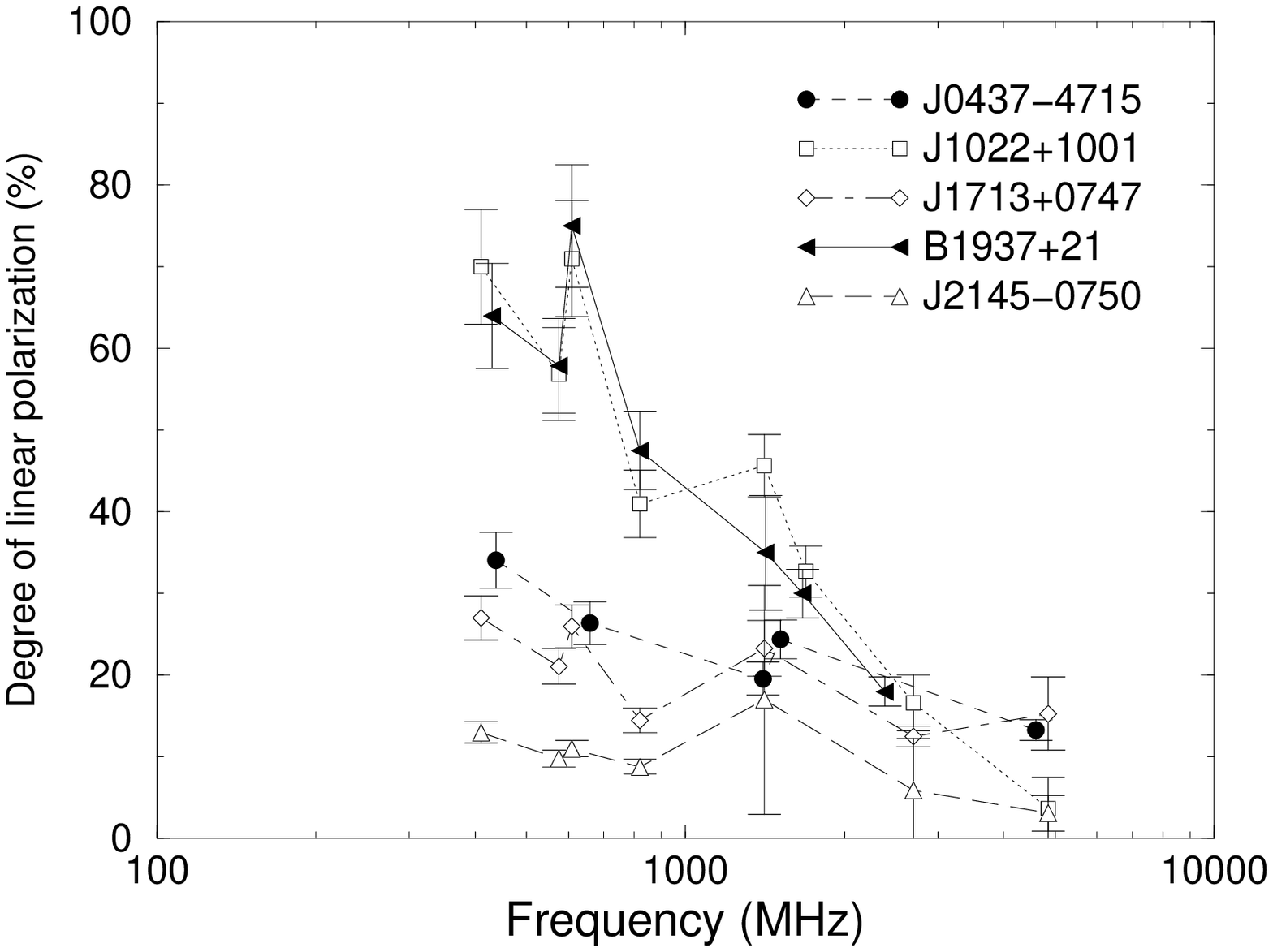}


\plotone{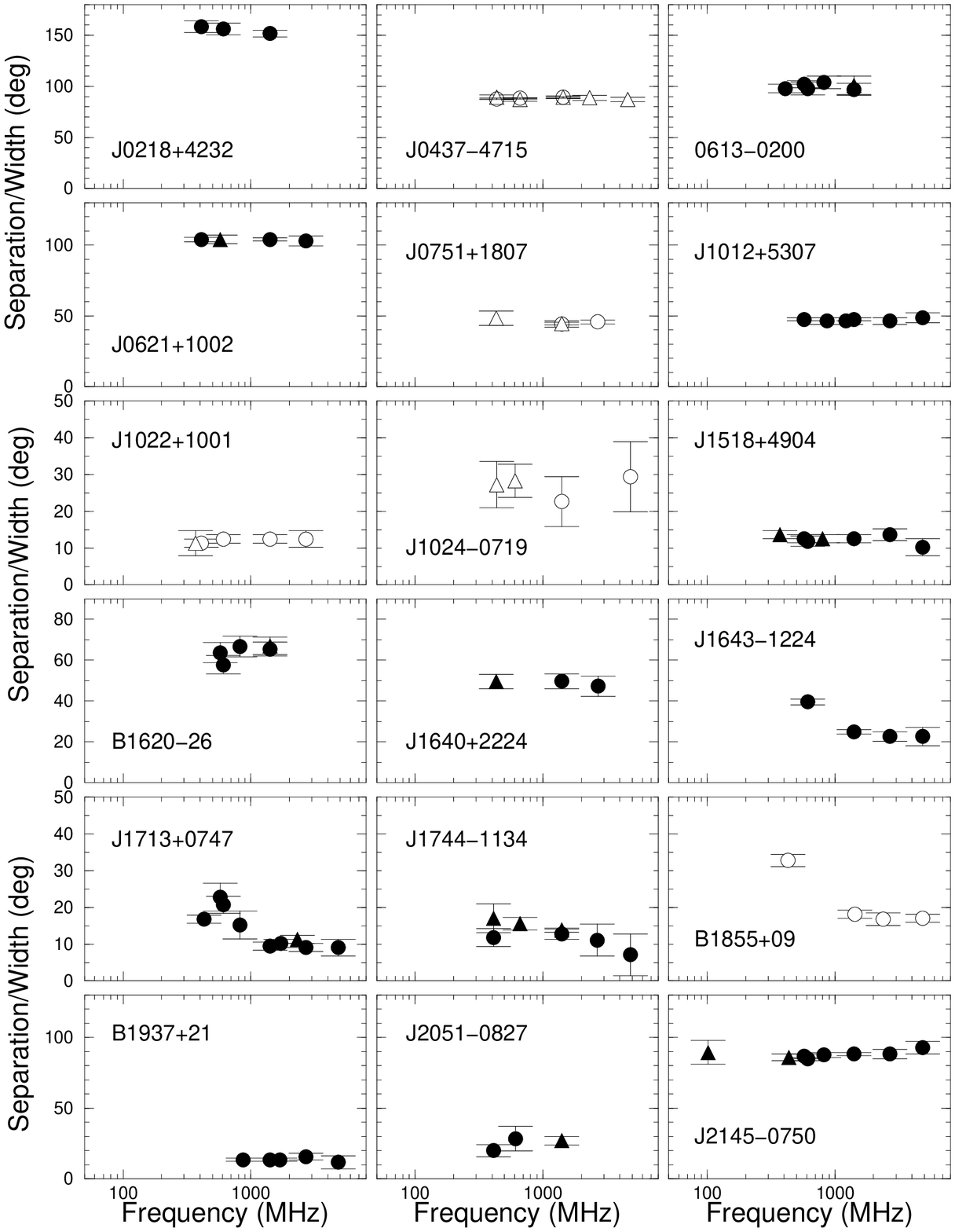}


\plotone{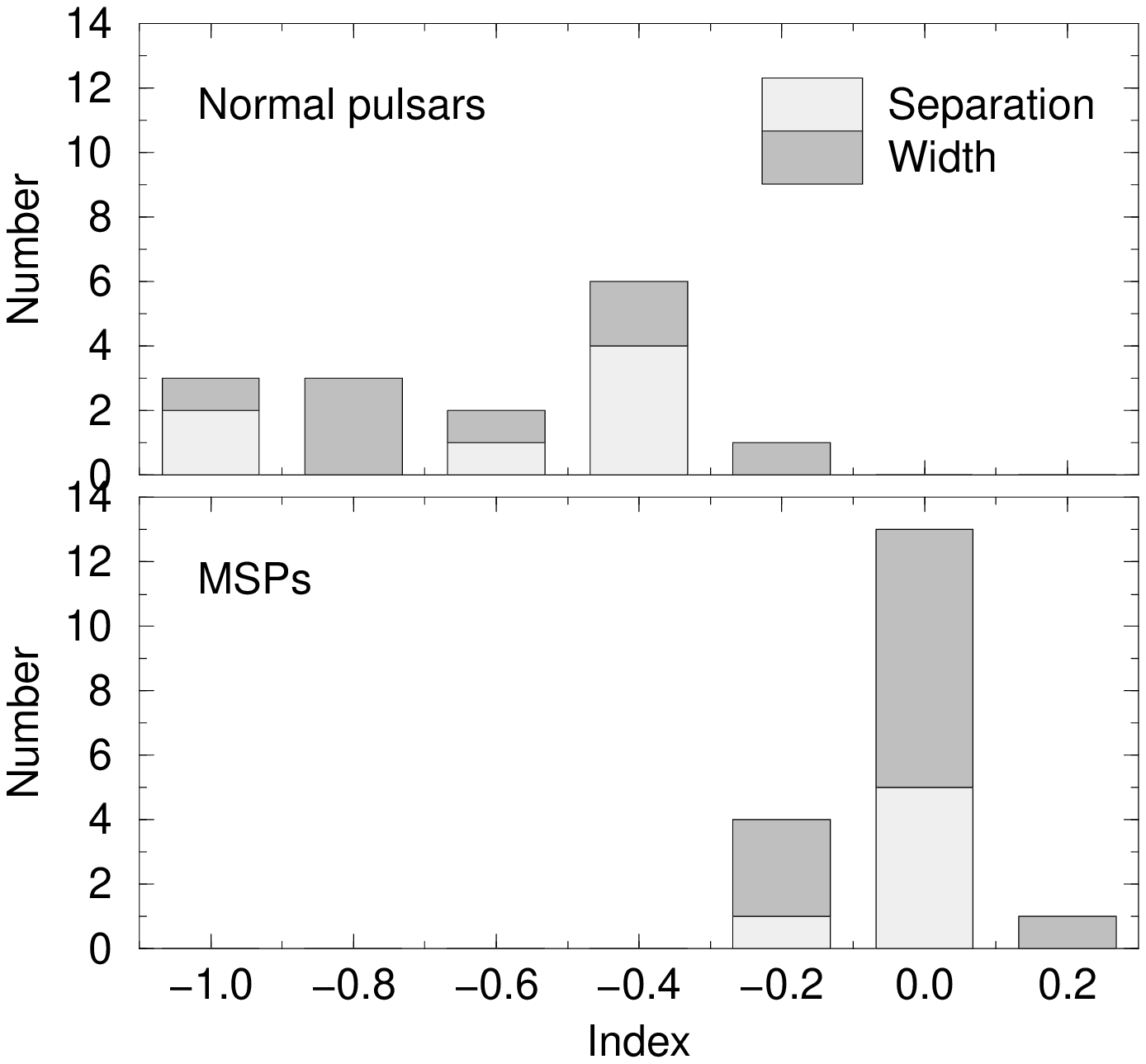}

\end{document}

%% file: tab1.tex


\begin{deluxetable}{lrcc}
\small
\tablenum{1}
\tablewidth{0pt}
\tablecaption{\label{oldobs} 
Previous observations of millisecond pulsars above 2 GHz.}
\tablehead{
\colhead{PSR} & \colhead{Period} & \colhead{Frequency} & \colhead{Ref.} \\
 \colhead{}      & \colhead{(ms)} & \colhead{(GHz)} & \colhead{}  }
\startdata
J0437$-$4715 & 5.757 & 2.3 & 1 \nl
             &       & 4.6 & 2 \nl
J0613$-$0200 & 3.062 & 2.3 & 1 \nl
J1022+1001 &  16.453 & 2.3 & 3 \nl
             &       & 4.9 & 4 \nl
J1045$-$4509 & 7.474 & 2.3 & 1 \nl
J1643$-$1224 & 4.621 & 2.3 & 1 \nl
J1713+0747   & 4.570 & 2.3 & 5 \nl
             &       & 4.9 & 4 \nl
B1855+09     & 5.362 & 2.3 & 6 \nl
             &       & 3.0 & 7 \nl
             &       & 4.9 & 4 \nl
B1937+21    & 1.558  & 2.3 & 7 \nl
             &       & 3.0 & 8 \nl
             &       & 4.8 & 9 \nl
             &       & 8.5 & 9 \nl
J2145$-$0750 & 16.052 & 4.9 & 4 \nl
\enddata
\tablerefs{1 - \cite{bbm+97}, 2 - \cite{mj95}, 3 - \cite{cam95a}, 
4 - \cite{kkwj97}, 5 - \cite{fwc93}, 
6 - \cite{ts90}, 7 - \cite{sti83}, 8 - \cite{ffb91}, 9 - \cite{ryb91}
}
\end{deluxetable}


%% file: tab2.tex
\begin{deluxetable}{lrcccc}
\small
\tablenum{2}
\tablewidth{0pt}
\tablecaption{\label{obstab} 
Observations of millisecond pulsars presented in this work.}
\tablehead{
\colhead{PSR} & \colhead{Period} & \colhead{Frequency} & \colhead{Backend} & 
\colhead{Bandwidth} & \colhead{Resolution\tablenotemark{a}} \\
 \colhead{}      & \colhead{(ms)} & \colhead{(GHz)} & \colhead{} &
\colhead{(MHz)} & \colhead{($\mu$s)} }
\startdata
J0218+4232 &   2.323 & 1.4 & EBPP\tablenotemark{b} & \phn45 & 143\phn \nl
J0621+1002 &  28.853 & 1.4 & EBPP & \phn56 & 113\phn \nl
           &         & 2.7 & EBPP & \phn80\tablenotemark{d} & 197\phn \nl
J0751+1807 &   3.479 & 2.7 & EBPP & \phn80 & 31.9 \nl
J1012+5307 &   5.256 & 0.9 & EBPP & \phn56 & 36.0 \nl
           &         & 1.2 & EBPP & \phn56 & 36.0 \nl
           &         & 1.4 & EBPP & \phn90 & 15.4 \nl
           &         & 2.7 & EBPP & \phn80 & 46.0 \nl
           &         & 4.8 & EBPP & 112 & 77.0 \nl
J1022+1001 &  16.453 & 1.4 & EBPP & \phn28  & 32.1 \nl
           &         & 2.7 & EBPP & \phn80 & 64.3 \nl
           &         & 4.8 & EBPP & 112 & 96.4 \nl
J1024$-$0719 & 5.162 & 1.4 & EBPP & \phn45  & 55.5 \nl
           &         & 4.8 & EBPP & 112 & 40.3  \nl
J1518+4904 &  40.935 & 1.4 & EBPP & \phn28 & 80.0 \nl
           &         & 2.7 & EBPP & \phn80 & 160 \nl
           &         & 4.8 & EBPP & 112 & 160 \nl
J1640+2224 &   3.163 & 2.7 & EBPP & \phn80 & 27.3 \nl
           &         & 4.8 & EBPP & 112 & 38.6 \nl
J1643$-$1224 & 4.621 & 1.4 & EBPP & \phn23  & 18.0 \nl
           &         & 2.7 & EBPP & \phn90  & 22.6 \nl
           &         & 4.8 & EBPP & 112 & 27.1 \nl
J1713+0747 &   4.570 & 1.4 & EBPP & \phn56 & \phn8.9 \nl
           &         & 1.7 & EBPP & \phn45 & \phn8.9 \nl
           &         & 2.7 & EBPP & \phn80 & 17.9 \nl
           &         & 4.9 & EPOS\tablenotemark{c} & \phn80 & 93.0 \nl
J1744$-$1134 & 4.075 & 1.4 & EBPP & 112 & 15.9 \nl
           &         & 2.7 & EBPP & \phn80 & 31.8 \nl
           &         & 4.8 & EBPP & 112 & 31.8 \nl
B1855+09   &   5.362 & 1.4 & EBPP & \phn28  & 10.5 \nl
           &         & 4.9 & EPOS & \phn80 & 77.4 \nl
B1937+21   &  1.558  & 0.9 & EBPP & \phn28 & \phn1.5 \nl
           &         & 1.4 & EBPP & \phn28 & \phn1.5  \nl
           &         & 1.7 & EBPP & \phn28 & \phn3.0 \nl
           &         & 2.7 & EBPP & \phn80 & \phn6.1 \nl
           &         & 4.8 & EBPP & 112 & 12.1 \nl
J2145$-$0750 & 16.052 & 1.4 & EBPP & \phn90 & 31.4 \nl
           &         & 2.7 & EBPP & 112 & 62.7 \nl
           &         & 4.9 & EPOS & 500 & 329.1\phn \nl
\enddata
\tablenotetext{a}{final resolution after data reduction}
\tablenotetext{b}{coherently de-dispersed}
\tablenotetext{c}{incoherently de-dispersed}
\tablenotetext{d}{limited due to terrestrial interference}
\end{deluxetable}

%% file: tab3.tex


\begin{deluxetable}{lrr@{$\; \pm \;$}lr@{$\; \pm \;$}lr@{$\; \pm \;$}lr@{$\; \pm \;$}l}
\tablenum{3}
\tablewidth{0pt}
\tablecaption{\label{spectab} 
Flux densities and spectral indices for millisecond pulsars at 2.7 and
4.9 GHz. The difference in spectral indices derived for separate power
laws below and above 1.4 GHz, $\Delta \alpha\equiv \alpha_{\le
1.4\mbox{\scriptsize GHz}} - \alpha_{\ge 1.4\mbox{\scriptsize GHz}}$
is quoted in the last column. See text for details.
}
\tablehead{
\colhead{PSR} & \colhead{P} & 
   \multicolumn{2}{c}{$S_{2695}$} & \multicolumn{2}{c}{$S_{4850}$} & 
 \multicolumn{2}{c}{spectral index} & \multicolumn{2}{c}{diff.~index}   \\
 \colhead{}      & \colhead{(ms)} & \multicolumn{2}{c}{(mJy)} & 
   \multicolumn{2}{c}{(mJy)} & \multicolumn{2}{c}{$\alpha$}  & 
   \multicolumn{2}{c}{$\Delta\alpha$}  }
\startdata
J0437$-$4715 &   5.757 & \multicolumn{2}{c}{...} & 8.5 & 2.5$^a$ & -1.17 & 0.06 & 1.4 & 0.3 \\
J0621+1002 &  28.853 & 0.3\phn & 0.2\phn  & \multicolumn{2}{c}{...} & $-1.9$\phn & 0.3\phn & \multicolumn{2}{c}{...$^b$}\nl
J0751+1807 &   3.479 & 0.8\phn & 0.3\phn  & \multicolumn{2}{c}{...} & $-1.2$\phn & 0.2\phn & 0.2 & 0.8$^c$\nl
J1012+5307 &   5.256 & 1.2\phn & 0.2\phn  & 0.2\phd & 0.1     & $-2.0$\phn & 0.1\phn & 0.1 & 0.3 \nl 
J1022+1001 &  16.453 & 1.9\phn & 0.3\phn  & 0.53    & 0.06     & $-1.5$\phn & 0.1\phn & 0.0 & 0.5 \nl 
\tablevspace{5pt}
J1024$-$0719 & 5.162 & \multicolumn{2}{c}{...} & 0.4\phn & 0.2\phn & $-1.4$\phn & 0.1\phn & \multicolumn{2}{c}{...$^d$} \nl
J1518+4904 &  40.935 & 0.8\phn & 0.2\phn  & 0.24    & 0.07     & $-1.5$\phn & 0.2\phn & 1.5 & $0.8^e$ \nl
B1620$-$26 &  11.076 & 0.2\phn & 0.1\phn  & \multicolumn{2}{c}{...} & $-1.7\phn$ & 0.1\phn & 1.5 & 0.9$^c$ \nl
B1640$+$2224 & 3.163 & 0.4\phn & 0.2\phn  & \multicolumn{2}{c}{...} & $-0.8$\phn & 0.3\phn  & \multicolumn{2}{c}{...$^f$} \nl
J1643$-$1224 & 4.621 & 1.4\phn & 0.5\phn  & 0.4\phn & 0.2\phn  & $-1.82$ & 0.05 & 0.5 & 0.5 \nl
\tablevspace{5pt}
J1713+0747 &   4.570 & 2.7\phn & 0.9\phn  & 0.8\phd & 0.2     & $-1.2$\phd & 0.2 & 0.6 & 0.4 \nl
J1744$-$1134 & 4.075 & 0.7\phn & 0.4\phn  & 0.19    & 0.06     & $-1.9$\phn & 0.1\phn & 0.4 & 0.4 \nl
B1855+09   &   5.362 & 2.3\phn & 0.7\phn  & 1.12    & 0.09     & $-1.33$ & 0.07 & 0.2 & 0.3 \nl
B1937+21   &  1.558  & 2.0\phn & 0.4\phn  & 1.0\phn & 0.2\phn  & $-2.13$ & 0.09 & 0.1 & 0.1 \nl
J2051$-$0827 & 4.509 & 0.3\phn & 0.2\phn  & \multicolumn{2}{c}{...} & $-1.5$\phn & 0.1\phn & 1.0 & 1.1$^c$ \nl
\tablevspace{5pt}
J2145$-$0750 & 16.052 & 2.2\phn & 0.5\phn & 0.4\phn & 0.1\phn  &$-1.88$ & 0.08 & 0.5 & 0.4 \nl
\enddata
\tablenotetext{a}{Measured by Manchester \& Johnston (1995) at 4.6 GHz. 
Uncertainty estimated to be 30\%.}
\tablenotetext{b}{No flux densities at 1.4 and 4.9 GHz.}
\tablenotetext{c}{No flux densities at 4.9 GHz. Quoted value only 
based on 1.4 and 2.7 GHz measurements.}
\tablenotetext{d}{Formal fit suggests a flatter spectrum at high
frequencies.}
\tablenotetext{e}{Possible turn-over around 400 MHz.}
\tablenotetext{f}{Only flux measurements at 0.4, 1.4 and 2.7 GHz available.}
\end{deluxetable}


%% file: tab4.tex


\begin{deluxetable}{lrcr@{$\; \pm \;$}lcr@{$\; \pm \;$}l}
\tablenum{4}
\tablewidth{0pt}
\tablecaption{\label{upperlim} 
Upper limits on flux densities for millisecond
pulsars not detected at 2.7 and 4.9 GHz. Values are calculated for
a $5\sigma$ detection of that pulse feature which is most prominent at
1.41 GHz (see Paper I).}
\tablehead{
\colhead{PSR} & \colhead{P} & \colhead{$S_{2695}^{upper}$} &
   \multicolumn{2}{c}{$S_{2695}^{exp}$} & \colhead{$S_{4850}^{upper}$} &
   \multicolumn{2}{c}{$S_{4850}^{exp}$} \\
 \colhead{}      & \colhead{(ms)} & \colhead{(mJy)} & \multicolumn{2}{c}{(mJy)} & 
   \colhead{(mJy)} & \multicolumn{2}{c}{(mJy)}  }
\startdata
J0613$-$0200 & 3.062 & 0.46 & 1.08 & 0.60       & 0.40 & 0.52\phn & 0.31\phn \nl
J0621+1002 &  28.853 & ... &  \multicolumn{2}{c}{...} & 0.32 & 0.07\phn & 0.30\phn \nl
J0751+1807 &   3.479 & ... &  \multicolumn{2}{c}{...} & 0.35 & 0.40\phn & 1.00\phn \nl
B1257+12   &   6.219 &  0.38 & 0.32 & 1.00 & ... & \multicolumn{2}{c}{...} \nl
B1534+12   &  37.904 &  0.05 & 0.09 & 0.04      & 0.04 & 0.015 & 0.007 \nl
\tablevspace{5pt}
B1640$+$2224 & 3.163 & ... & \multicolumn{2}{c}{...}  & 0.44 & 0.42\phn & 1.10\phn \nl
J2051$-$0827 & 4.509 & ... & \multicolumn{2}{c}{...}  & 0.44 & 0.3\phn\phn & 0.4\phn\phn \nl
\enddata
\end{deluxetable}


%% file: tab5.tex


\begin{deluxetable}{lcccr@{$\; \pm \;$}lr@{$\; \pm \;$}l}
\tablenum{5}
\tablewidth{0pt}
\tablecaption{\label{polobstab} 
Summary of polarization observations.  We quote observing frequency,
bandwidth, effective resolution and the degree of linear, $\Pi_L$,
 and total polarization, $\Pi_T$.}
\tablehead{
\colhead{PSR} & \colhead{$\nu$} & \colhead{BW} & 
 \colhead{$\Delta t_{\rm eff}$} &
   \multicolumn{2}{c}{$\Pi_L \equiv \frac{<\;L\;>}{<\;I\;>}$} & 
   \multicolumn{2}{c}{$\Pi_T \equiv \frac{<\;\sqrt{L^2+V^2}\;>}{<\;I\;>}$ } \\
 \colhead{} & \colhead{(GHz)} & \colhead{(MHz)} & \colhead{($\mu$s)} 
 & \multicolumn{2}{c}{(\%)} &  \multicolumn{2}{c}{(\%)} }
\startdata
J1022+1001 & 1.69 & \phd28 & 16.1 & 32.7 & 3.1 & 34.0 & 3.1 \\
           & 2.69 & \phd40 & 174 & 16.6 & 3.4 & 18.0 & 3.1  \\
           & 4.85 & \phd40 & 33.9 & \phd3.7 & 3.8 & \phd6.5 & 5.5 \\
\tablevspace{5pt}
J1713+0747 & 2.69 & \phd40 & \phd6.4 & 12.5 & \phd1.3 & 12.7 & 1.3 \\
           & 4.85 & \phd80 & 93.1 & 15.3 & \phd4.5 & 16.7 & 6.0  \\
\tablevspace{5pt}
J2145$-$0750 & 2.69 & \phd80 & 304 & \phd5.9 & 6.4 & 12.3 & 5.1 \\
             & 4.85 & 200 & 132 & \phd3.1 & 2.2 & 7.5 & 5.5 \\
\enddata
\end{deluxetable}


%% file: tab6.tex


\begin{deluxetable}{llr@{$\; \pm \;$}lr@{$\; \pm \;$}lr@{$\; \pm \;$}l}
\tablenum{6}
\tablewidth{0pt}
\tablecaption{\label{tabsepwidth} 
Frequency dependences and values at 1 GHz 
of component separation (C), and profile width
(W), respectively, and de-polarization index for MSPs with measured 
polarization at 5 GHz (see text for details).}
\tablehead{
 & \multicolumn{3}{c}{Measure} & \multicolumn{2}{c}{} & \multicolumn{2}{c}{} \\
\colhead{PSR} & \colhead{Type} & \multicolumn{2}{c}{at 1 GHz (deg)} & 
 \multicolumn{2}{c}{$\gamma$} & \multicolumn{2}{c}{$\epsilon$} }
\startdata
J0218+4232   & W$^a$ & 153 & 3\phd & $-0.04$  & 0.03 & \multicolumn{2}{c}{...} \\
J0437$-$4715 & C$^b$ &  88.7 & 0.3 &  0.00    & 0.01 & $-0.38$ & 0.05 \\
J0613$-$0200 & W$^c$ & 101 & 3\phd &  0.00    & 0.05 & \multicolumn{2}{c}{...} \\
J0621+1002   & C$^d$ & 104.0 & 0.8 &  0.00    & 0.01 & \multicolumn{2}{c}{...} \\
J0751+1807   & C$^{d,e}$ & 45 & 1\phd & $-0.02$   & 0.05 &  \multicolumn{2}{c}{...} \\
J1012+5307   & W$^a$ & 47.3 & 0.8 &  0.00    & 0.02 &  \multicolumn{2}{c}{...} \\
J1022+1001   & C$^f$ & 12.3 & 0.7 &  0.05    & 0.08 & $-0.56$ & 0.07 \\
J1024$-$0719 & C$^g$ & 27.3 & 0.3 &  0.0\phd & 0.1\phd & \multicolumn{2}{c}{...} \\
J1518+4904   & W$^h$ & 12.6 & 0.5 & $-0.03$  & 0.06 & \multicolumn{2}{c}{...} \\
B1620$-$26   & W$^{c,d}$ & \phd65 & 2\phd &  0.08    & 0.08 & \multicolumn{2}{c}{...} \\
J1640+2224   & W$^c$ & \phd49& 2\phd & $-0.02$  & 0.06 & \multicolumn{2}{c}{...} \\
J1643$-$1224 & W$^i$ & 31.3 & 0.8 & $-0.27$  & 0.05 & \multicolumn{2}{c}{...} \\
J1713+0747   & W$^{i,j}$ & 13.5 & 0.5 & $-0.26$  & 0.05 & $-0.34$ & 0.06 \\
J1744$-$1134 & W & 14.0 & 0.6 & $-0.2$\phd & 0.1\phd & \multicolumn{2}{c}{...} \\
B1855+09     & C$^k$ & 23.8 & 0.8 & $-0.30$  & 0.03 & \multicolumn{2}{c}{...} \\
B1937+21     & W$^a$ & 13.5 & 0.8 & 0.0\phd  & 0.1\phd & $-0.76$ & 0.07 \\
             & C$^l$ & 172.6 & 0.3 & 0.00  & 0.01 &  \multicolumn{2}{c}{} \\
J2051$-$0827 & W$^m$ & 26 & 2\phd & 0.2\phd & 0.2 & \multicolumn{2}{c}{...} \\
J2145$-$0750 & W & 87.3 & 0.6 & 0.02 & 0.01 & $-0.5$ & 0.2 \\
\enddata
\footnotesize
\tablenotetext{a}{Main pulse feature.}
\tablenotetext{b}{Using strongest outer components at $-40^\circ$ and 
$+50^\circ$ longitude. Additional data taken from Navarro et al.~(1997).}
\tablenotetext{c}{Additional data from Sallmen (1998) and
Stairs et al.~(1999).}
\tablenotetext{d}{Using strong outer components.}
\tablenotetext{e}{Additional data taken from Lundgren et al.~(1995),
Sallmen (1998) and Paper I.}
\tablenotetext{f}{Using prominent leading and trailing component, see 
Kramer et al.~(1999).}
\tablenotetext{g}{Using components remaining at 4.9 GHz.}
\tablenotetext{h}{Width referring to intensity of components remaining
at 4.9 GHz.}
\tablenotetext{i}{Width measurements probably influenced by unresolved component
at low frequencies.}
\tablenotetext{j}{Additional data at 1.4 GHz taken from Sallmen (1998).}
\tablenotetext{k}{Using components as resolved at 4.9 GHz.}
\tablenotetext{l}{Using separation between peaks of main and interpulse.} 
\tablenotetext{m}{Additional data from Paper I and Stairs et al.~(1999).}
\end{deluxetable}
